\documentclass[twocolumn]{aastex701}
\usepackage{amsmath}
\usepackage[utf8]{inputenc}

\newcommand{\vect}[1]{\boldsymbol{#1}}

\newcommand{\dif}{\mathrm{d}}
\newcommand{\e}{\mathrm{e}}
\newcommand{\yr}{\,\mathrm{yr}}
\newcommand{\Myr}{\,\mathrm{Myr}}

\newcommand{\AU}{\,\mathrm{AU}}

\newcommand{\MSol}{M_{\odot}}

\newcommand{\kms}{\mathrm{km \, s^{-1}}}

\newcommand{\xhat}{\vect{\hat{x}}}
\newcommand{\yhat}{\vect{\hat{y}}}
\newcommand{\zhat}{\vect{\hat{z}}}

\newcommand{\gaia}{\textit{Gaia}}
\newcommand{\ciera}{Center for Interdisciplinary Exploration and Research in Astrophysics (CIERA), Northwestern University, 1800 Sherman Ave., Evanston, IL 60201, USA}

\shorttitle{The fate of \gaia's wide binaries}
\shortauthors{C.\ E.\ O'Connor}

\journalinfo{}
\received{2025 September 2}
\submitjournal{ApJ}

\begin{document}

\title{The fate of \gaia's wide binaries: Interplay of white-dwarf recoil and tidal capture}

\author[0000-0003-3987-3776]{Christopher E.\ O'Connor}
\affiliation{\ciera}
\email[show]{christopher.oconnor@northwestern.edu}

\begin{abstract}
    White dwarfs (WDs) receive natal velocity boosts of $\sim 1 \, \kms$ due to recoil from asymmetric mass loss during the late asymptotic giant branch (AGB) stage. 
    In a wide binary, the acceleration of a proto-WD exerts a torque, modifying the orbital eccentricity. 
    Potential signatures of this effect have been detected in \gaia's astrometric binary sample. 
    At the same time, an AGB star's puffy envelope facilitates strong tidal interactions in binaries with periapsis separations of a few AU, 
    capturing the companion into a tighter orbit and potentially driving the system towards a common-envelope phase.  
    Using an analytical model for wide binary evolution under asymmetric AGB mass loss, 
    we find that recoil can induce tidal interactions in up to $30\%$ of initially wide binaries on the AGB or post-AGB 
    for initial separations of $\sim 100 \mbox{--} 1000 \AU$. 
    We relate these interactions to three recent observational puzzles: 
    (i) The dearth of wide WD+MS and WD+WD binaries in \gaia\ DR3 with eccentricities $\gtrsim 0.9$. 
    (ii) The formation of moderately eccentric WD+MS and WD+WD binaries with orbital periods of $\sim 100 \mbox{--} 1000$ days, 
    which may happen via a high-eccentricity common-envelope phase. 
    (iii) The origin of low-luminosity, long-timescale, dust-obscured transients 
    towards AGB progenitors, such as the ongoing event WNTR23bzdiq in M31.
    Our findings have potential implications for the survival and dynamical evolution of planetary systems around WD progenitors, to be investigated in future works. 
\end{abstract}


\section{Introduction}

With the rapid maturation of time-domain and multi-messenger astronomy, 
it is more important than ever to understand the interplay of stellar evolution and orbital dynamics across the entire stellar mass domain. 
The late evolutionary stages of low- and intermediate-mass stars ($\sim 1 \mbox{--} 8 \MSol$), the progenitors of white dwarfs (WDs), 
harbor theoretical uncertainties that have long hindered efforts to develop reliable prescriptions for binary stellar evolution. 

Two of the most important and least certain ingredients in binary evolutionary models are tidal interactions and stellar mass loss. 
On the main sequence (MS), tides drive gradual orbital evolution in close binaries by synchronizing stellar spins and circularizing the orbit. 
When one component departs the main sequence (MS) and accordingly swells in size, it becomes more susceptible to strong tides, 
up to and including a contact phase or merger through Roche lobe overflow. 
Meanwhile, mass loss determines the nature of stellar remnants formed in a given system; 
it also counteracts tidal orbital decay and, if mass is expelled asymmetrically, modifies the orbit's shape and orientation. 
Roughly speaking, the relative importance of these two processes is determined by binary separation: 
Tidal interactions drop off drastically in strength with increasing separation, 
becoming effectively negligible beyond $\sim 10 \AU$ for even the largest stellar radii. 
Mass loss is potentially important at any separation, 
but its effect on the orbit may be qualitatively different between close and wide binaries, depending on the nature of mass loss. 

Arguably the most impactful data product to date in the study of low- and intermediate-mass binary evolution 
is the (recently decommissioned) \gaia\ mission's catalog of astrometric binaries in the Galactic field \citep{ElBadry2021}. 
\gaia's snapshot of orbital and physical parameters of Galactic binaries 
enables an unprecedented level of insight into the underlying physics of these processes and their astrophysical implications. 

AGB mass loss takes place via dense, dust-driven winds with typical mass-loss rates $|\dot{m}|$ up to a few $10^{-5} \MSol \yr^{-1}$ \citep[e.g.,][]{Reimers1975_mdots, Renzini1981_AGBwinds, Decin2019_AGBmdots}. 
Significant progress has occurred in recent years in understanding the physics of these winds, 
but a comprehensive, predictive theory of AGB mass loss remains elusive
\citep[e.g.,][]{HoefnerOlofsson2018_AGBmasslossreview, Decin2021_AGBmasslossreview}.  
Various lines of evidence indicate that at least some AGB stars produce asymmetric winds, 
imparting appreciable velocity boosts to newborn WDs, 
known as WD natal kicks or recoil.\footnote{We prefer the latter term, since ``kick'' may inappropriately connote an instantaneous acceleration.}  
Young WDs in core-collapse globular clusters are observed to have a less concentrated radial distribution than older WDs within the same clusters. 
This can be explained if WDs receive boosts of $\Delta V \sim 1 \, \kms$ over a timescale no greater than the cluster crossing time, some $10^{5} \yr$ \citep[e.g.,][]{Heyl2007, Davis+2008, Fregeau+2009}. 
More recently, \gaia\ has revealed a drop-off in the occurrence rates of WD+MS and WD+WD binaries with separations greater than $\sim 10^{3} \AU$ \citep{ElBadry2018}. 
Again, this can be explained by $\sim 1 \, \kms$ boosts imparted on a timescale shorter than the Keplerian orbital period at $\sim 10^{3} \AU$, which is also $\sim 10^{5} \yr$ for typical progenitor masses. 
Thus, WD recoil accelerations are much smaller and probably more gradual than the natal kicks imparted to neutron stars and black holes by supernovae. 

Previous studies of the consequences of WD recoil have, for the most part, considered considered the impulsive limit of the problem, as with supernova kicks \citep[e.g.,][]{Fregeau+2009, Izzard+2010, ElBadry2018}. 
However, the impulse approximation requires an acceleration timescale much shorter than the orbital period. 
In binaries with semi-major axes less than $\sim 10^{3} \AU$, we expect the effects of recoil to be described better under the adiabatic approximation. 
This distinction is analogous to the difference between impulsive and low-thrust maneuvers in spaceflight; 
indeed, the recoil acceleration $\vect{g}$ of a mass-losing star may be described using the classical rocket equation: 
\begin{equation} \label{eq:rocket_equation}
    \vect{g} = - \frac{\dot{m} \vect{w}}{m},
\end{equation}
where $m$ is the stellar mass and $\vect{w}$ is the effective 1D exhaust velocity of ejecta relative to the star. 

The difference between impulsive and gradual recoil implies qualitatively different outcomes for binary systems that experience WD recoil depending on their separation. 
Specifically, the dissolution of binaries by WD recoil described by \citet{ElBadry2018} should not occur in binaries initially closer than $\sim 10^{3} \AU$, even if $\Delta V$ exceeds the binary's escape velocity. 
This is because the acceleration is distributed across the orbit, zeroing out the net change in the binary's orbital energy. 
Instead, the binary's eccentricity and orientation undergo a slow, potentially large-amplitude oscillation, 
the characteristics of which depend on the magnitude and orientation of the recoil acceleration \citep[e.g.,][]{Heyl2007_wdkickbinaries, HwangZakamska2025}. 
Many other studies have considered variations on the rocket-effect problem in different astrophysical settings 
\citep[e.g.,][]{Rasio1992_pulsarplanet, Lai2001_pulsarrocket, Agalianou2023_pulsarrocket, Hirai2024}. 

In this paper, we study the fate of \gaia's wide binary systems 
under the interplay of stellar mass loss and tidal interactions.
Our goals are as follows: 
First, to provide an elementary description of the dynamics 
in the appropriate regime for a majority (though by no means all) of \gaia's binary systems. 
Second, to demonstrate clearly how \gaia's characterization of wide binary eccentricities 
can refine constraints on specific aspects of AGB mass loss (see also \citealt{HwangZakamska2025}). 
Third, to explore connections with questions of current interest in the areas of binary evolution and time-domain astronomy.  
In Section \ref{s:binary_kick_model}, we present a simple derivation of the secular evolution of binaries under WD recoil 
and discuss how the process can drive wide binaries towards collisions. 
In Section \ref{s:gaia_binaries}, we use this model to perform rudimentary population synthesis experiments for \gaia's wide binaries with semi-major axes $10 \AU \lesssim a \lesssim 10^{3} \AU$. 
We predict how the orbital distribution function of \gaia's wide binaries differs 
between the MS+MS, WD+MS, and WD+WD samples as a result of WD recoil and induced tidal interactions between binary components.  
In Section \ref{s:discussion}, we discuss implications of our results for the origin of AU-scale WD binaries in the \gaia\ catalog and the emerging class of optical and infrared transients associated with giant stars. 
In Section \ref{s:conclusion}, we summarize our main conclusions and recommend potential refinements for future work. 

\section{Evolution of a binary under WD recoil} \label{s:binary_kick_model}

\subsection{Dynamical model}

Consider a Keplerian binary system in which the components undergo adiabatic mass variations with recoil. 
The dynamics of such a system has been studied previously by \citet{Heyl2007_wdkickbinaries}. 
We present a simplified analysis of the problem in the secular (i.e., orbit-averaged) limit. 

We make the following assumptions: 
\begin{enumerate}
    \item The components of a wide binary or planetary system are point-like for purposes of determining their instantaneous gravitational attraction. 
    
    \item Recoil imparts a small relative acceleration $\vect{g}(t)$ 
    in a fixed direction with respect to inertial space 
    according to equation (\ref{eq:rocket_equation}).     
    Other interactions between the binary components and the stellar wind(s), 
    such as gravitational torques, drag forces, and accretion, are negligible. 
    
    \item There exists a critical separation $r_{c}$ such that, if the periapsis separation of the binary falls below this value, 
    strong tidal interactions rapidly reduce the orbital period and eccentricity, 
    removing the system from the \gaia\ wide-binary sample. 
    We refer to this as ``tidal capture.''  
\end{enumerate}
The dynamics of the system during the directional phase of mass loss is described 
in the effective one-body framework by the following Hamiltonian function \citep[e.g.,][]{Hadjidemetriou1963, Hadjidemetriou1966, Heyl2007_wdkickbinaries, Hirai2024}: 
\begin{align} \label{eq:general_Ham}
    H &= \frac{1}{2} \vect{v}^{2} - \frac{G M(t)}{r} - \vect{g}(t) \cdot \vect{r},
\end{align}
where $M(t) = m_{1}(t) + m_{2}(t)$ is the system's total mass, with $m_{1}$ and $m_{2}$ the individual component masses. 
The reduced mass, $\mu(t) = m_{1}(t) m_{2}(t) / M(t)$, does not appear in the equations of motion.
The vectors $\vect{r}$ and $\vect{v}$ give the binary's relative position and velocity in the rest frame of $m_{1}$. 
This Hamiltonian defines the gravitational Stark problem (or accelerated Kepler problem), whose solutions are well known \citep[e.g.,][]{Namouni2006, Namouni2007, Heyl2007_wdkickbinaries, LantoineRussell2011, Hirai2024}. 
Approximate analytical solutions based on perturbation theory are also possible when the parameters are time-dependent. 
In systems with directional mass loss, equation (\ref{eq:rocket_equation}) may be generalized to include rocket-like accelerations for both stars: 
\begin{equation}
    \vect{g}(t) = - \left( \frac{\dot{m}_{1}}{m_{1}} \vect{w}_{1} - \frac{\dot{m}_{2}}{m_{2}} \vect{w}_{2} \right).
\end{equation}
We simplify this expression by considering systems in which mass loss occurs for only one star at a time; 
without loss of generality, we let the subscript `1' refer to the mass-losing star for the remainder of this discussion.  

We denote the mean Keplerian orbital elements as follows: 
semi-major axis $a$, eccentricity $e$, inclination $I$, longitude of the ascending node $\Omega$, argument of the periapsis $\omega$, and mean anomaly $l$. 
It is also convenient to define the eccentricity vector and reduced angular momentum vector: 
\begin{widetext}
\begin{subequations}
\begin{align}
    \vect{e} &= e \left[ \zhat \sin\omega \sin{I} + \xhat \left( \cos{\Omega} \cos{\omega} - \sin{\Omega} \sin{\omega} \cos{I} \right) + \yhat \left( \sin{\Omega} \cos{\omega} + \cos{\Omega} \sin{\omega} \cos{I} \right) \right], \\
    \vect{j} &= \left( 1 - e^{2} \right)^{1/2} \left[ \zhat \cos{I} + \sin{I} \left( \xhat \sin{\Omega} - \yhat \cos{\Omega} \right) \right],
\end{align}
\end{subequations}
\end{widetext}
where the unit vectors $(\xhat, \yhat, \zhat)$ form a right-handed triad in the frame of $m_{1}$. 
These vectors describe the shape and orientation of the orbit. 

Since the recoil acceleration is small compared to the Newtonian gravitational interaction, 
we may average over the instantaneous Keplerian orbit to obtain the Hamiltonian describing the system's secular evolution: 
\begin{equation} \label{eq:secular_Ham}
    \langle H \rangle \equiv \frac{1}{2 \pi} \int_{0}^{2 \pi} H \, \dif l = - \frac{G M}{2 a} - \frac{3}{2} a \vect{g} \cdot \vect{e}. 
\end{equation}
Since mass loss and recoil are adiabatic effects by hypothesis, 
the principle of adiabatic invariance implies that 
the action variable conjugate to $l$, $L = (G M a)^{1/2}$, is conserved. 
Thus, the semi-major axis evolves in inverse proportion to the system's total mass:
\begin{equation} \label{eq:a(t)_secular}
    a(t) = a_{i} \left[ \frac{m_{1i} + m_{2i}}{m_{1}(t) + m_{2}(t)} \right],
\end{equation}
where the subscript $i$ indicates the initial value of a quantity. 
This is a well-known result for isotropic mass loss; 
we conclude that it also holds for asymmetric mass loss in the adiabatic limit. 
The secular equations of motion arising from equation (\ref{eq:secular_Ham}) take a concise form in terms of $\vect{e}$ and $\vect{j}$ \citep[e.g.,][]{Milkanovitch1939, Tremaine2023_dynplsys}: 
\begin{subequations} \label{eq:Milankovitch_eqns}
\begin{align}
    \frac{\dif \vect{e}}{\dif t} &= - \frac{1}{L} \left( \vect{j} \times \frac{\partial}{\partial \vect{e}} + \vect{e} \times \frac{\partial}{\partial \vect{j}} \right) \langle H \rangle = \vect{\gamma} \times \vect{j}, \\
    \frac{\dif \vect{j}}{\dif t} &= - \frac{1}{L} \left( \vect{j} \times \frac{\partial}{\partial \vect{j}} + \vect{e} \times \frac{\partial}{\partial \vect{e}} \right) \langle H \rangle = \vect{\gamma} \times \vect{e},
\end{align}
\end{subequations}
where 
\begin{equation} \label{eq:def_omega_g}
    \vect{\gamma} = \frac{3}{2} \left( \frac{a}{G M} \right)^{1/2} \vect{g}
\end{equation}
is a vector describing the frequency and orientation of the orbital modulation.  
Although we assume that $\vect{g}$ maintains a fixed direction in inertial space, 
equations (\ref{eq:secular_Ham}), (\ref{eq:Milankovitch_eqns}), and (\ref{eq:def_omega_g}) remain valid when relaxing this assumption, 
provided the directional variation also be adiabatic. 

It is straightforward to decouple equations (\ref{eq:Milankovitch_eqns}) by taking the time derivative of the first line and substituting the second (or vice versa). 
We then have
\begin{equation}
    \frac{\dif^{2} \vect{j}}{\dif t^{2}} = \vect{\gamma} \times \left( \vect{\gamma} \times \vect{j} \right)
\end{equation}
and likewise for $\vect{e}$. 
If we express each vector in terms of its components parallel (subscript $\parallel$) and perpendicular (subscript $\perp$) to $\vect{g}$, 
then we obtain 
\begin{equation}
    \frac{\dif^{2} \vect{j}_{\perp}}{\dif t^{2}} = - \gamma^{2} \vect{j}_{\perp}, \ \ \vect{j}_{\parallel} = {\rm const.},
\end{equation}
and likewise for $\vect{e}_{\perp}$ and $\vect{e}_{\parallel}$.
Thus, $\vect{e}$ and $\vect{j}$ undergo a harmonic oscillation 
with frequency $\gamma$ in the plane perpendicular to $\vect{g}$. 
In practice, the secular evolution of a real system may not be a true harmonic oscillation 
because the magnitude and direction of $\vect{\gamma}$ can vary in real systems. 
In Appendix \ref{app:stark_general_solution}, we provide the general solution for $\vect{e}(t)$ and $\vect{j}(t)$ when $\vect{g}$ is fixed in direction but varies adiabatically in magnitude via equation (\ref{eq:rocket_equation}). 

\subsubsection{Special case: circular orbit, constant frequency}

For purposes of illustration, let us momentarily take $\gamma$ to be constant 
and consider an initially circular orbit inclined by an angle $I_{0}$ to the recoil acceleration. 
Without loss of generality, we may construct the coordinate system with $\vect{g}$ along the positive $z$-axis 
and the ascending node on the positive $y$-axis (i.e., $\Omega_{0} = \pi/2$); 
then 
\begin{subequations}
\begin{align}
    \vect{e}(t) &= - \yhat \sin{I_{0}} \sin{\gamma t}, \\
   \vect{j}(t) &= \zhat \cos{I_{0}} + \xhat \sin{I_{0}} \cos{\gamma t},
\end{align}
\end{subequations}
The eccentricity and inclination at time $t$ are
\begin{subequations}
\begin{align}
    e(t) &= |\sin{I_{0}}| \left[ \frac{1 - \cos{2 \gamma t}}{2} \right]^{1/2}, \\ 
    \cos{I(t)} &= \frac{\cos{I_{0}}}{[ \cos^{2}{I_{0}} + \sin^{2}{I_{0}} \cos^{2}(\gamma t) ]^{1/2}} \nonumber \\
    &= \cos{I_{0}} \left[ \cos^{2}{I_{0}} + \frac{\sin^{2}{I_{0}} \left( 1 + \cos{2 \gamma t} \right)}{2} \right]^{-1/2}.
\end{align}
\end{subequations}
Thus we see that the scalar eccentricity and inclination oscillate with a fundamental frequency $2 \gamma$, with a half-period (i.e., the time elapsed between consecutive minima and maxima) of
\begin{widetext}
\begin{subequations} \label{eq:def_TStark}
\begin{align}
    T_{\rm Stark} &= \frac{\pi}{4 \gamma} \\
    &= 2.2 \times 10^{5} \yr \left( \frac{g}{10 \, \kms \Myr^{-1}} \right)^{-1} \left( \frac{a}{100 \AU} \right)^{-1/2} \left( \frac{M}{2 \MSol} \right)^{1/2} \\
    &\sim 10^{5} \yr \left( \frac{\tau_{\rm accel}}{10^{5} \yr} \right) \left( \frac{V}{1 \, \kms} \right)^{-1} \left( \frac{a}{100 \AU} \right)^{-1/2} \left( \frac{M}{2 \MSol} \right)^{1/2},
\end{align}
\end{subequations}
\end{widetext}
where on the third line we express the average acceleration of a proto-WD as $g = \Delta V / \tau_{\rm accel}$. 
The maximum eccentricity achieved by an initially circular binary is $|\sin{I_{0}}|$. 
For $I_{0} = 0$ or $\pi$, the orbit-averaged torque vanishes. 
For $I_{0} = \pi/2$, the maximum eccentricity is formally unity; 
in practice, short-range forces or collisions enforce a maximum eccentricity slightly less than 1 \citep[e.g.,][]{Liu+2015}. 

Equation (\ref{eq:def_TStark}) neatly confirms the intuitive expectation that WD recoil can modify the orbit significantly when $\Delta V \gtrsim (G M / a)^{1/2}$. 
This recalls the well-known result that an instantaneous kick $\Delta V$ applied to an initially circular orbiter 
changes the eccentricity by $\Delta e \sim V / v_{\rm orb}$, unbinding the system ($\Delta e \gtrsim 1$) when $\Delta V \gtrsim v_{\rm orb}$ \citep[e.g.,][]{Akiba+2024}. 
However, the secular approximation guarantees that the system remains formally bound even if the acceleration continued indefinitely, 
since the total orbital energy can never become positive through adiabatic mass loss. 
Instead, we expect systems with $\Delta V \gtrsim v_{\rm orb}$ to be more vulnerable to tidal capture, 
since for favorable recoil directions their minimum periapsis distances may approach the postulated critical value $r_{c} \ll a$. 
For typical values $M \sim 1 \mbox{--} 8 \MSol$ and $\Delta V \sim 1 \, \kms$, 
we expect systems with initial separations $a_{i} \gtrsim 10 \AU$ 
to experience appreciable eccentricity modulation due to WD recoil. 

We have verified that our analytical model gives a suitably accurate approximation for our purposes
by comparing its predictions to direct integrations of binaries with {\tt REBOUND}/{\tt REBOUNDx} \citep{ReinLiu2012_rebound, Tamayo2020_reboundx}. 
We describe this exercise in Appendix \ref{app:validate}. 
In brief, we find that the secular approximation yields acceptably accurate predictions 
for binaries whose semi-major axes remain below $\sim 10^{3} \AU$. 
Additionally, the secular model predicts the minimum periapsis separation $r_{p} = a(1-e)$ accurately, 
even when the approximation breaks down for the individual orbital elements $a$ and $e$. 
This has bearing on the question of tidal capture in highly eccentric systems (Section \ref{s:binary_kick_model:tidal_capture}).

\subsection{Model for AGB mass loss and recoil} \label{s:binary_kick_model:AGB}

We adopt the following simplified model for AGB mass loss: 
We assume that the outflow from an AGB star $m_{k}$ is initially spherically symmetric and undergoes a transition to an asymmetric configuration at late times. 
During the latter stage, the newborn WD receives a gradual acceleration $\vect{g}_{k}$ in a fixed direction in inertial space. 
By virtue of working under the adiabatic approximation, we need not specify the mass loss rate $\dot{m}_{k}$ and acceleration $\vect{g}_{k}$ explicitly as functions of time. 
It is sufficient to know the following quantities:
\begin{itemize}
    \item The star's initial mass (denoted $m_{ki} = m_{k,{\rm MS}}$), its mass at the beginning of the recoil stage ($m_{ka}$), and its final mass ($m_{kf} = m_{k,{\rm WD}}$). 
    The MS and WD masses are connected by a theoretical or empirical initial--final mass relation. 
    The second is a free parameter in our model. To control this quantity in our numerical experiments, we define a dimensionless parameter $0 < f_{a} \leq 1$ such that
    \begin{equation} \label{eq:define_fa}
        m_{ka} = f_{a} m_{ki} + (1-f_{a}) m_{kf}.
    \end{equation}
    To be precise, $f_{a}$ gives the fraction of the AGB star's envelope mass that remains at the onset of recoil; 
    the limiting cases $f_{a} = 1$ and $f_{a} \simeq 0$ correspond to recoil occurring throughout the AGB phase and in the final moments before becoming a WD, respectively. 
    
    \item The total velocity boost $\Delta V_{k} = \int \vect{g}_{k}(t') \, \dif t'$, which is related to the effective exhaust velocity $\vect{w}_{k}$ via 
    \begin{equation} \label{eq:define_DeltaV}
        \Delta \vect{V}_{k} = -\vect{w}_{k} \ln\left( \frac{m_{ka}}{m_{kf}} \right),
    \end{equation}
    with $m_{ka}$ given by equation (\ref{eq:define_fa}). 
\end{itemize}

\subsection{Tidal capture in eccentric binaries} \label{s:binary_kick_model:tidal_capture}

It is of interest to track the minimum periapsis distance $r_{p} = a(1-e)$ achieved by a binary during its post-MS evolution. 
If $r_{p}$ is of the order of the AGB star's maximum radius, then strong tidal interactions may occur during closest approach. 
The outcomes of these interactions are unclear and sensitive to the detailed stellar structure, mass loss history, and orbital evolution. 
The possibilities range from partial tidal circularization of the orbit to physical contact between the binary components via Roche lobe overflow; 
the latter could potentially initiate a common-envelope stage. 
We refer to these outcomes collectively as tidal capture. 

To study this possibility, we introduce an additional parameter to our model, denoted $r_{c}$, representing a critical separation within which tidal capture occurs. 
The order of magnitude of $r_{c}$ is given by the periapsis separation for which the AGB star fills its Roche lobe \citep[e.g.,][]{Eggleton1983}:
\begin{equation}
    r_{c} = R_{1} \frac{0.6 q^{2/3} + \ln(1 + q^{1/3})}{0.49 q^{2/3}}, 
\end{equation}
where $R_{1}$ is the AGB stellar radius and $q$ the mass ratio. 
Using values $0.1 \leq q \leq 1$ yields $2.6 \leq r_{c}/R_{1} \leq 4.8$. 
Thus, for typical $R_{1} \sim 1 \AU$, values of $r_{c}$ up to $\sim 5 \AU$ are reasonable. 

\subsection{Example}

\begin{figure}
    \centering
    \includegraphics[width=\columnwidth]{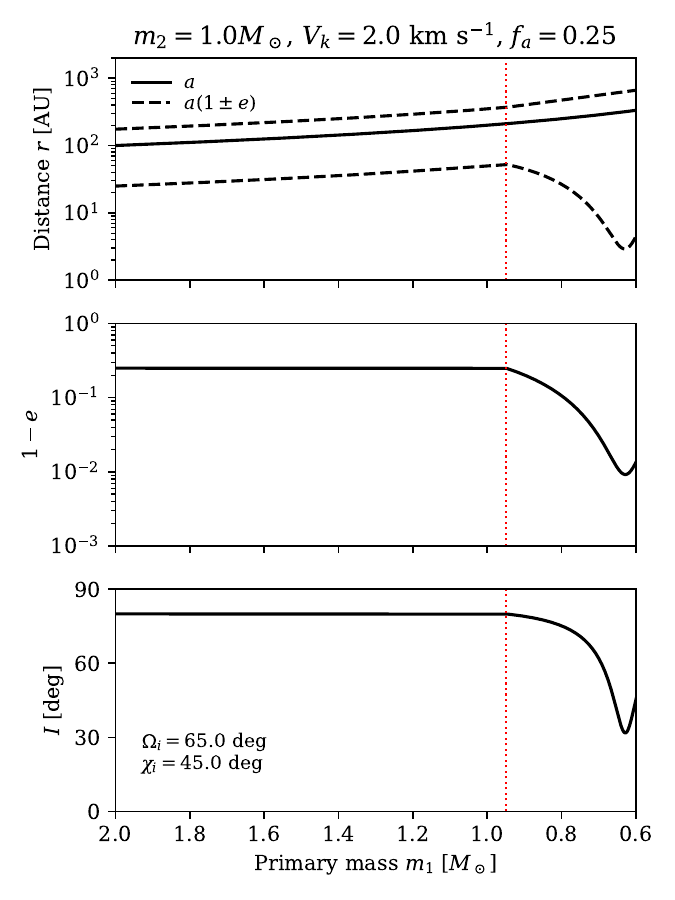}
    \caption{Orbital evolution of a typical wide binary as the primary star undergoes AGB mass loss with recoil. 
    The top panel shows the semi-major axis $a$ (solid curve) and the periapsis and apoapsis separations $a(1 \pm e)$ (dashed). 
    The middle panel shows the eccentricity as $1-e$ with a logarithmic vertical scale. 
    The bottom panel shows the orbital inclination. 
    The binary components have initial masses of $m_{1i} = 2.0 \MSol$ and $m_{2} = 1.0 \MSol$; 
    the initial orbital elements are $a_{i} = 100 \AU$, $e_{i} = 0.75$, $I_{i} = 80 \deg$, $\Omega_{i} = 165 \deg$, and $\omega_{i} = 45 \deg$. 
    The primary undergoes recoil with a total velocity boost of $V_{k} = 2.0 \, \kms$ along the z-axis. 
    Recoil is initiated at a mass given by equation (\ref{eq:define_fa}) with $f_{a} = 0.25$ (red vertical lines).}
    \label{fig:analytic_Ex1}
\end{figure}

Figure \ref{fig:analytic_Ex1} shows the evolution of a binary with MS masses of $m_{1i} = 2.0 \AU$ and $m_{2} = 1.0 \MSol$. 
The primary star evolves into a $0.6 \MSol$ WD, 
receiving a boost of $V_{k} = 2.0 \, \kms$ along the $z$-axis; 
the recoil stage begins at $m_{1a} = 0.95 \MSol$, corresponding to $f_{a} = 0.25$. 
The initial conditions are chosen such that the binary achieves a maximum eccentricity of $\simeq 0.99$, 
corresponding to a minimum periapsis of $\simeq 2.9 \AU$. 
Hence, this system could potentially be vulnerable to tidal capture, depending on the assumed $r_{c}$.

It is impossible to determine the initial conditions and the mass-loss and recoil parameters for any given WD binary. 
Thus, it is only in a statistical sense that our model can be used to reconstruct or forecast the evolution of individual systems. 
However, by examining the evolution of \emph{ensembles} of simulated binary systems and comparing their properties to those of the observed \gaia\ sample, 
it may be possible to place constraints on the magnitude and directional distribution of WD recoil accelerations. 
This motivates the exercises carried out in the next section. 

\section{Evolution of \gaia'\lowercase{s} wide binaries} \label{s:gaia_binaries}

The \gaia\ binary sample, through a combination of well-characterized completeness, curation for high-quality kinematic data, and sheer size, 
has produced a wealth of insights on wide binary formation and dynamical evolution 
(see the recent review by \citealt{ElBadry2024}). 
A particularly intriguing discovery is that wide binaries in the Galactic field exhibit a separation-dependent eccentricity distribution. 
Although precise eccentricity measurements for individual binaries in the catalog are difficult, 
it is nonetheless possible to characterize the underlying eccentricity distribution via the technique of \citet{Hwang2022_binecc}. 
Assuming that wide binaries at a given separation $s$ follow a power-law eccentricity distribution $f_{e}(e) = (1+\alpha) e^{\alpha}$, 
these authors reported that $\alpha$ increases significantly with separation: 
specifically, they found a smooth transition from a near-uniform distribution ($\alpha \approx 0$) at $s \lesssim 10^{2} \AU$ 
to a superthermal distribution ($\alpha \approx 1.3$) at $s \gtrsim 10^{3} \AU$. 
Numerous subsequent studies have explored potential explanations for the origin of this trend, 
including interstellar turbulence in star-forming regions \citep{Xu+2023, Mathew+2024}, 
perturbations from the Galactic tidal field and stellar flybys \citep{Hamilton2022, ModakHamilton2023, HamiltonModak2024}, 
and dynamical interactions in star clusters \citep{RoznerPerets2023, GinatPerets2024, Atallah+2024}. 
Whatever its origin, the observed relation between binary separation and eccentricity turns out to be highly useful for our purposes, 
in that it constrains the initial conditions for models of post-MS dynamical evolution. 
We will show that, if the eccentricities of wide MS+MS binaries obey the trend reported by \citet{Hwang2022_binecc}, 
the eccentricity distributions of WD+MS and WD+WD binaries inevitably differ from the MS+MS population, 
and from each other, in a manner determined by the nature of AGB mass loss 
and the efficiency of tidal capture in highly eccentric AGB binaries. 

\begin{figure*}
    \centering
    \includegraphics[width=\textwidth]{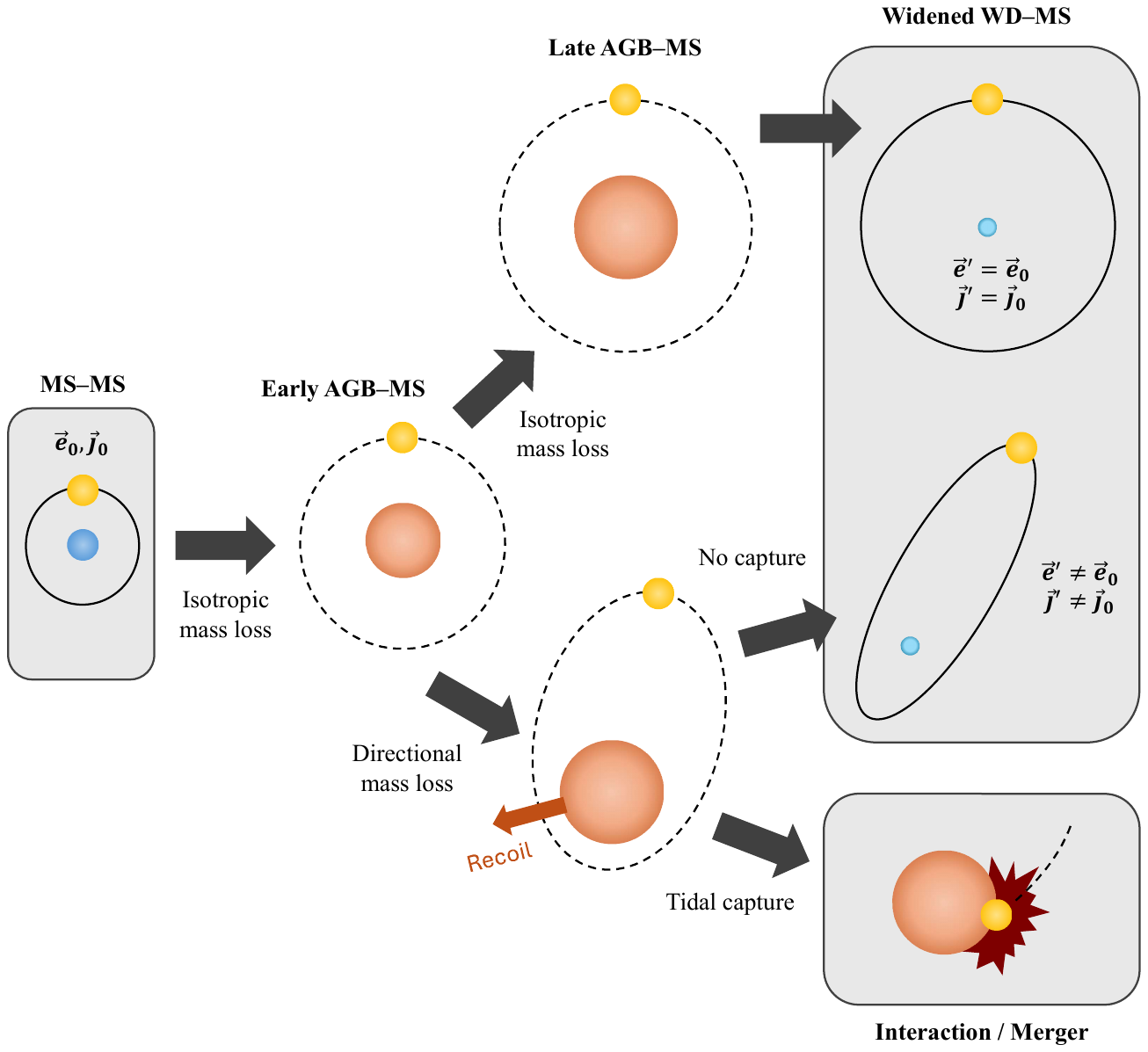}
    \caption{Illustration of possible evolutionary pathways of a wide binary during the post-MS stages of the primary star, 
    highlighting the expected outcomes with or without directional mass loss in the late AGB phase. 
    If the periapsis separation of the binary grows sufficiently small due to recoil, 
    then tidal capture may occur, leading to an eccentric binary interaction or merger. 
    If the secondary star is massive enough to become a WD within the age of the Galaxy, 
    then the processes shown above will repeat. 
    }
    \label{fig:cartoon}
\end{figure*}

Figure \ref{fig:cartoon} illustrates the potential evolutionary pathways for a wide MS+MS binary, 
focusing on the potential outcomes of the primary star's AGB phase. 
We assume that an initially wide MS+MS binary undergoes isotropic mass loss during its post-MS evolution until reaching the late AGB phase. 
At this point, different possible fates branch off depending on whether the star undergoes recoil 
and, if so, whether its periapsis distance becomes small enough for tidal capture to occur. 
In any case, there are two possible outcomes, both of observational interest: 
widened MS--WD (or WD+WD) binaries, 
whose eccentricities may or may not have been modified by recoil; 
and eccentric interacting AGB binaries, the evolutionary fates of which are unclear. 
A surviving WD+MS binary may experience a second round of evolution 
if the secondary MS star is massive enough to evolve within a Hubble time. 

\subsection{Procedure}

Here we describe our procedure for modeling the post-MS dynamical evolution of \gaia\ wide binaries. 
In brief, we generated ensembles of mock MS+MS binary systems broadly mimicking the \gaia\ sample in terms of orbital distribution function. 
We then employed the analytical model described in Section \ref{s:binary_kick_model} to compute the post-MS evolution of these same binaries, 
thereby obtaining ensembles of WD+MS and WD+WD binaries potentially observable by \gaia. 
We compare the eccentricity distributions of these populations at each evolutionary stage to illustrate the likely signatures of WD recoil and recoil-induced tidal capture. 

\subsubsection{Initial conditions and input physics}

We consider binary systems in which at least one member becomes a WD within the age of the Galaxy, modeling the post-MS evolution of each system as follows. 
We considered all stars above $0.95 \MSol$ to be potential WD progenitors and disregarded the evolution of all stars below this mass. 
We convert initial masses into WD masses using the empirical relation of \citet{ElBadry2018_InitialFinalMassRelation}, 
a piecewise-linear function on the interval $0.95 \leq m_{i}/\MSol \leq 8.0$. 

We assume that AGB winds are initially isotropic but transition to being somewhat directional later on. 
The transition is assumed to occur when the stellar mass reaches the value given by equation (\ref{eq:define_fa}), using a default value of $f_{a} = 0.1$ unless otherwise specified. 
Since our dynamical model employs the adiabatic approximation for stellar mass loss, 
we do not need to specify a particular function form for the stellar mass-loss rate $\dot{M}(t)$ and acceleration $\vect{g}(t)$. 
The final state is completely determined by the initial and final stellar masses and the vector $\Delta \vect{V}_{k}$ (equation \ref{eq:define_DeltaV}). 
We draw the magnitude of $\Delta \vect{V}_{k}$ from a Maxwellian distribution with dispersion $\sigma_{V}$ 
and its direction from a spherically isotropic distribution \citep[e.g.,][]{ElBadry2018}. 
We assume that this distribution is the same for all stars. 
Given these quantities, we integrate equations (\ref{eq:Milankovitch_eqns}) as described in Appendix \ref{app:stark_general_solution}. 
This yields analytic expressions for the orbital elements in the WD+MS or WD+WD stage.

For each simulated system, we track the minimum periapsis distance $r_{p} = a(1-e)$ achieved during the AGB phase. 
If this value is less than or equal to the critical separation $r_{c}$, 
we assumed that tidal capture occurred and removed the binary from consideration. 
As with $f_{a}$ and $\sigma_{V}$, we use the same value of $r_{c}$ for all stars in a given ensemble. 
A reasonable value for $r_{c}$ can be estimated as the periapsis separation for which the AGB star fills its Roche lobe,
using the approximate analytical formula due to \citet{Eggleton1983}: 
\begin{equation}
    r_{c} = R_{1} \frac{0.6 q^{2/3} + \ln(1 + q^{1/3})}{0.49 q^{2/3}}, 
\end{equation}
where $R_{1}$ is the AGB stellar radius and $q$ the mass ratio. 
This formula gives $r_{c} \approx 2.6 R_{1}$ ($4.8 R_{1}$) for $q = 1$ ($0.1$). 
Since AGB stars have typical radii of the order of an AU, representative values of $r_{c}$ should be of the order of a few AU. 
We therefore considered ensembles in which the critical distance takes one of the values $\{ 0, 1, 2, 5 \} \AU$ for all systems. 
Ensembles with $r_{c} = 0$ examine the evolution if tidal capture were negligible. 
Those with finite, larger values suffer attrition from tidal capture, 
which is reflected in the eccentricity distributions of WD+MS and WD+WD binaries. 

\subsubsection{Modeling the final eccentricity distribution}

As described above, we are less interested in the eccentricity evolution 
of individual binary systems than the evolution of the overall eccentricity distribution as a function of separation. 
We must therefore adopt a fiducial model for the distribution 
that can adequately and concisely describe the full range of population-synthesis outcomes. 
We find that the simple power-law model adopted by \citet{Hwang2022_binecc} 
for the eccentricity distribution of wide MS+MS binary stars 
is inadequate to describe the distribution for post-MS binaries. 
The main issue is the turnover at high eccentricities in ensembles where tidal capture is important. 
We find that a beta distribution provides an improved description for our purposes. 
This distribution has support on $0 \leq e \leq 1$ 
with a probability density function given by
\begin{equation} \label{eq:beta_distribution}
    f_{e}(e) = \frac{\Gamma(\alpha+\beta+2)}{\Gamma(\alpha + 1) \Gamma(\beta + 1)} e^{\alpha} (1 - e)^{\beta}, 
\end{equation} 
where $\Gamma$ is the gamma function. 
This model has been previously applied to the distribution of exoplanet eccentricities \citep[e.g.,][]{Kipping2013, VanEylen2019}. 
It main advantage in this case is that it naturally generalizes the power-law model of \citet{Hwang2022_binecc} 
while introducing only one additional free parameter. 
The interpretation of the parameter $\alpha$ is roughly the same as before, 
measuring the shape of the distribution at low eccentricities. 
For a fixed $\alpha > 0$, the value of $\beta$ determines the nature of the turnover at high eccentricities, 
with larger positive values corresponding to more pronounced turnovers. 
We expect $\beta \approx 0$ for distributions without an appreciable turnover, i.e.\ those consistent with a simple power law. 

Although the beta distribution is useful for our proof-of-concept study, 
we should note two main shortcomings of our approach. 
It is somewhat difficult to judge the minimum value of $\beta$ that constitutes an appreciable turnover. 
For purposes of illustration, we suppose \gaia\ astrometry can constrain $\alpha$ and $\beta$ 
with typical statistical uncertainties of $\sigma_{\alpha} = \sigma_{\beta} = 0.1$, 
comparable to those reported by \citet{Hwang2022_binecc, Hwang2022_twins} and \citet{HwangZakamska2025}. 
Additionally, the beta-distribution model assumes that $f_{e}(e)$ is nonzero and differentiable over the entire interval $0 < e < 1$. 
Consequently, it cannot accurately describe distributions with discontinuities, 
such as in Set~I-C below (Fig.\ \ref{fig:alphabeta_SetII}, third row). 
Models with more free parameters, such as a multi-step function \citep{Hwang2022_binecc}, may be more appropriate in such cases. 

\subsection{Idealized binary ensembles}

\begin{figure*}
    \centering
    \includegraphics[width=0.83\textwidth]{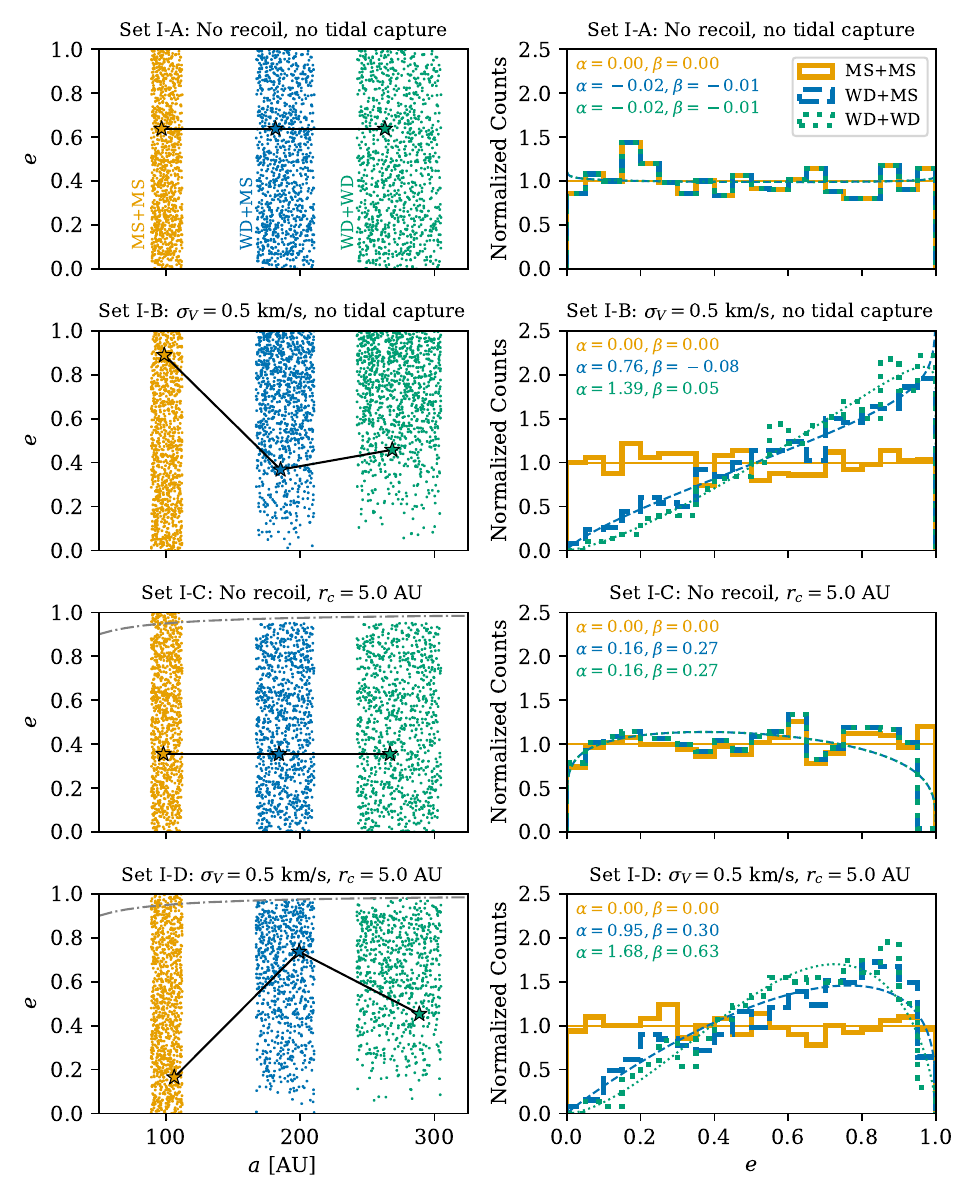}
    \caption{The predicted evolution of idealized wide binary populations 
    for different assumptions about dynamical evolution during the AGB stage: 
    no recoil and no tidal capture (Set I-A, top row); 
    recoil but no tidal capture (Set I-B, upper middle); 
    tidal capture but no recoil (Set I-C, lower middle); 
    and both recoil and tidal capture (Set I-D, bottom). 
    In the left-hand panels, small dots show the properties of individual systems during their MS+MS (orange), WD+MS (blue), and WD+WD stages (green). 
    A single system's progression between stages is highlighted as a series of filled stars connected by black lines. 
    In cases with tidal capture taken into account, the dot-dashed grey curve marks the critical periapsis distance $a (1-e) = r_{c}$.
    In the right-hand panels, thick lines show eccentricity histograms for all systems at each stage with bin widths of $0.05$. 
    Thin lines show the probability density functions for the known initial conditions (for MS+MS systems) 
    and the best-fitting beta distributions (for WD+MS and WD+WD systems), 
    with corresponding $\alpha$ and $\beta$ values listed in the upper-left-hand corner.
    The initial stellar masses are $m_{1} = 2 \MSol$ and $m_{2} = 1 \MSol$ in all systems. 
    The initial semi-major axes are distributed across the interval $90 \leq a/{\rm AU} \leq 110$, and the initial eccentricities are uniformly distributed.}
    \label{fig:orbit_distribution_example}
\end{figure*}

We begin by studying the evolution of an idealized ensemble of binary systems 
where we use the same pair of initial component masses ($2 \MSol + 1 \MSol$) for all systems.
We do this in two stages, first considering binaries within a narrow range of initial separations with a flat eccentricity distribution (Set I), 
and then binaries with the separation-dependent eccentricity distribution from the \gaia\ sample (Set II). 

\subsubsection{Set I}

Consider an ensemble of binaries whose components all have initial masses of $m_{1} = 2 \MSol$ and $m_{2} = 1 \MSol$, 
whose initial semi-major axes lie between $90$ and $110 \AU$, and whose eccentricities are drawn from a flat distribution ($\alpha = 0$). 
The binary components eventually become WDs with masses of $0.6 \MSol$ and $0.5 \MSol$, 
respectively \citep[e.g.,][]{ElBadry2018_InitialFinalMassRelation}. 
To isolate the effects of WD recoil and tidal capture, 
we compute the evolution of this ensemble in four cases: 
\begin{itemize}
    \item Case A: A `vanilla' model, where AGB mass loss is fully isotropic ($\sigma_{V} = 0$) and where tidal capture is neglected ($r_{c} = 0$). 
    \item Case B: A `recoil-only' model, where kicks occur with $f_{a} = 0.1$ and $\sigma_{V} = 0.5 \,\kms$ but $r_{c} = 0$.
    \item Case C: A `tidal-capture-only' model, where $r_{c} = 5 \AU$ but $\sigma_{V} = 0$.  
    \item Case D: A `full' model in which $f_{a} = 0.1$, $\sigma_{V} = 0.5 \,\kms$, and $r_{c} = 5 \AU$.
\end{itemize}

The panels in the left column of Figure \ref{fig:orbit_distribution_example} show the evolution of each ensemble in the $a$--$e$ plane, 
with the MS+MS, WD+MS, and WD+WD stages shown in different colors. 
In the right column, we show the incremental eccentricity distribution at each stage of evolution.

In Set I-A, systems undergo adiabatic orbital expansion at constant eccentricity, as expected. 
The marginalized eccentricity distribution is unchanged within a narrow range of semi-major axes. 
In other words, WD+MS and WD+WD binaries merely inherit the eccentricity distribution of their MS+MS progenitor systems, 
when one accounts for the adiabatic expansion of orbits. 
Even so, this would produce a distinct observational signature in the \gaia\ binary catalog, as we will show.

In Set I-B, recoil causes phase-mixing of the binaries' eccentricities (and orientations) on top of adiabatic orbital expansion. 
An individual system's eccentricity may increase or decrease, as shown in Figure \ref{fig:orbit_distribution_example}. 
However, the overall distribution of eccentricities changes its shape in a coherent manner, 
becoming increasingly skewed towards high eccentricities in the WD+MS and WD+WD stages.  
The distribution is consistent with a  power law with $\alpha \approx 0.8$ ($\alpha \approx 1.4$) for WD+MS (WD+WD) systems.\footnote{It is noteworthy that the combination of orbital expansion and WD recoil 
can convert a subthermal eccentricity distribution into a superthermal one. 
This cannot occur for purely conservative secular effects, 
such as Galactic tides \citep{Hamilton2022}.}

In Set I-C, stellar binaries whose initial periapsis separations are less than $r_{c}$ are lost to tidal capture. 
Those with larger separations survive with their eccentricities unchanged. 
This has the effect of truncating the eccentricity distribution for WD+MS and WD+WD systems 
at $e_{\rm max} \approx 1 - r_{c}/100 \AU \approx 0.95$, 
but otherwise preserving its original shape. 
Note that tidal capture events can occur only between the MS+MS and WD+MS stages in this experiment, 
since the surviving WD+MS systems have periapsis separations strictly greater than $r_{c}$. 
Moreover, because of orbital expansion between the WD+MS and WD+WD stages, 
the minimum periapsis separation is greater for WD+WD systems. 

Finally, in Set I-D, we see how WD recoil and tidal capture interact. 
As before, individual systems increase or decrease in eccentricity under WD recoil, 
but the distribution shifts towards higher eccentricities on average. 
Again, however, binaries that either begin with high eccentricities, 
or that reach high eccentricities through WD recoil, are lost. 
Thus, the WD+MS and WD+WD distributions are distinctly non-monotonic, 
being depleted in both quasi-circular ($e \lesssim 0.3$) and near-radial ($e \gtrsim 0.9$) orbits 
compared to the original population. 
The WD+WD distribution displays a moderately steeper slope at low eccentricities than the WD+MS distribution, 
due to the tidal capture of additional systems during the secondary AGB phase. 

\subsubsection{Set II} \label{s:gaia_binaries:setII}

\begin{figure*}
    \centering
    \includegraphics[width=0.95\textwidth]{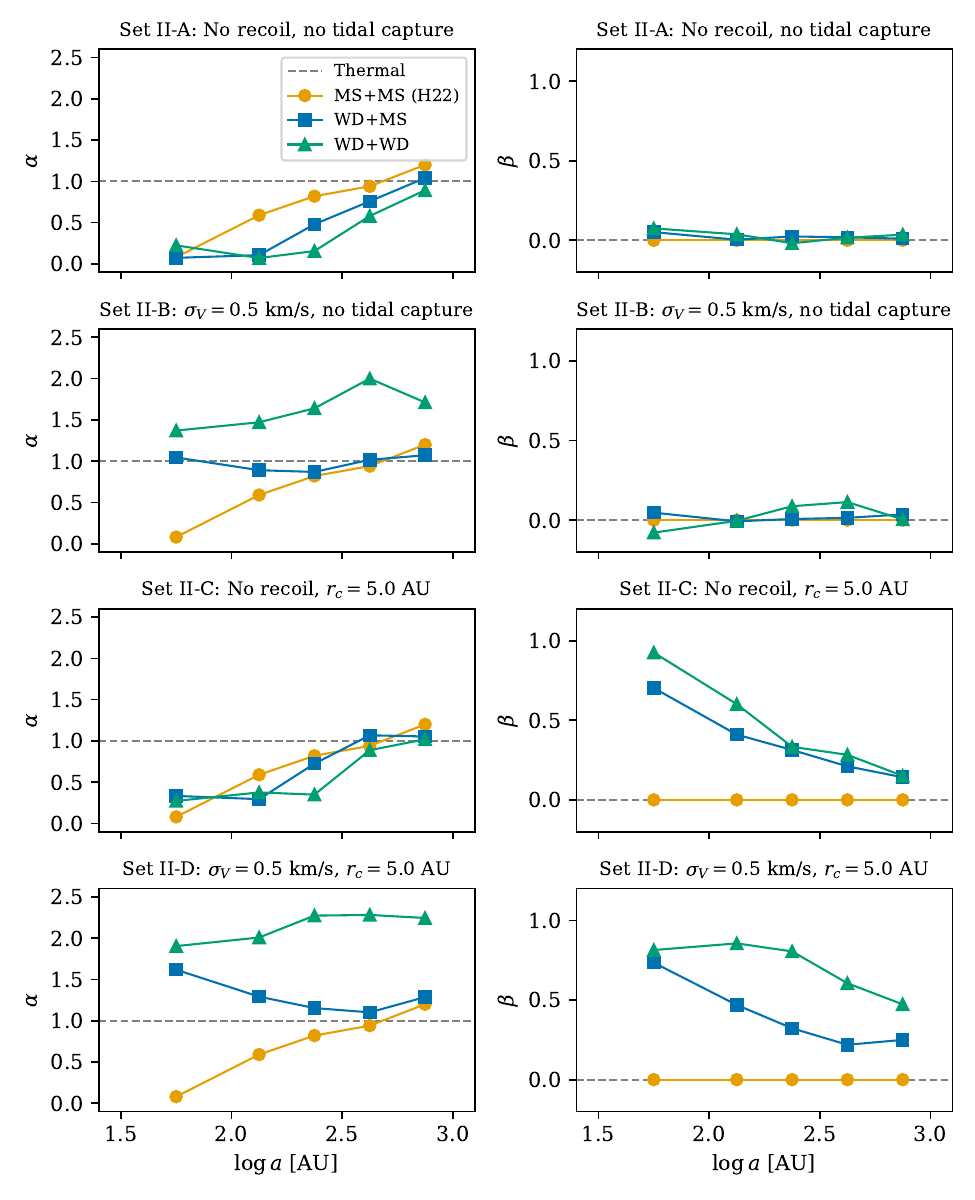}
    \caption{Predicted eccentricity distributions as a function of semi-major axis for evolving binary populations: 
    the MS+MS systems (orange circles) are mapped to the WD+MS (blue squares) and then WD+WD (green triangles) stages using our analytical model. 
    We fit a beta distribution (equation \ref{eq:beta_distribution}) to the eccentricities of the surviving systems at each stage. 
    The best-fitting $\alpha$ ($\beta$) values for each bin of semi-major axis are plotted in the left (right) column. 
    The horizontal grey lines indicate the values $\alpha = 1$ and $\beta = 0$, appropriate for a thermal distribution.}
    \label{fig:alphabeta_SetII}
\end{figure*}

We now consider ensembles of $2 \MSol + 1 \MSol$ binaries with a more realistic distribution of initial orbital elements 
inspired by \gaia\ astrometry. 
We draw initial semi-major axes from a log-uniform distribution from $30$ to $1000 \AU$ 
and initial eccentricities from a power-law distribution $f_{e}(e;a) = [1+\alpha(a)] e^{\alpha(a)}$, 
where $\alpha(a)$ is a piecewise function given by the best-fitting values in Table 1 of \citet{Hwang2022_binecc}. 
As before, we construct and evolve four ensembles under the same assumptions as Cases A through D above. 
This time, however, we focus on how the eccentricity distribution changes as a function of semi-major axis. 
At each evolutionary stage, we binned the binaries by current semi-major axis as follows:
$\log(a/{\rm AU}) \in [1.5,2.0)$, $[2.0,2.25)$, $[2.25,2.5)$, $[2.5,2.75)$, $[2.75, 3.0]$. 
These correspond to the separation bins used by \citet{Hwang2022_binecc} 
when characterizing the MS+MS eccentricity distribution. 
As before, we exclude systems that reach $a (1-e) \leq r_{c}$ or that have final $a > 10^{3} \AU$. 
Within each bin, we fit a beta distribution (equation \ref{eq:beta_distribution}) to the binary eccentricities using {\tt scipy} \citep{Virtanen2020_scipy}.

In Figure \ref{fig:alphabeta_SetII}, we show the best-fitting $\alpha$ and $\beta$ values as a function of $a$ 
for each binary ensemble at each evolutionary stage. 
For different assumptions about recoil and tidal capture, 
the trend of the fitted $\alpha$ and $\beta$ changes distinctly 
between the MS+MS, WD+MS, and WD+WD stages. 
From the top row of panels (Set II-A), we see that without WD recoil or tidal capture, 
WD+MS and WD+WD binaries inherit the eccentricity distribution for MS+MS systems at their original separation. 
The value of $\beta$ is consistent with zero (i.e.\ a pure power-law $f_{e} \propto e^{\alpha}$) within our fiducial error bars.
In the second row (Set II-B), we see the effect of adding WD recoil but neglecting tidal capture: 
The value of $\alpha$ typically increases at a given separation 
(WD+MS systems wider than $a \sim 10^{2.5} \AU$ being exceptions). 
Overall, the WD+MS binaries have roughly thermal eccentricities ($\alpha \approx 1$, $\beta \approx 0$) at all separations in this case. 
The WD+WD binaries, meanwhile, have superthermal eccentricities ($1.4 \lesssim \alpha \lesssim 2.0$, $\beta \approx 0$) at all separations. 
This goes to show how effectively even modest recoil velocities ($\lesssim 1 \, \kms$) can modify wide binary orbits. 
The third row (Set II-C) shows the effect of neglecting recoil 
while allowing tidal capture at small periapse separations. 
Without the phase-mixing induced by recoil, 
the $\alpha$ values at each stage are similar to those from Set II-A. 
However, due to the removal of binaries with small periapsis distances after the MS+MS stage, 
we obtain moderate-to-large $\beta$ values for WD+MS and WD+WD systems. 
The value of $\beta$ tends to decrease with increasing $a$; 
this is because wider binary systems can be 
more eccentric while avoiding tidal capture for a given $r_{c}$. 
Finally, in the bottom row (Set II-D), we see the  prediction 
for systems in which both recoil and tidal capture can occur. 
Unsurprisingly, we find the greatest differences between the three evolutionary stages in this case. 
The $\alpha$ values are broadly similar to those in Set II-B, 
and similarly the $\beta$ values resemble those in Set II-C. 
However, due to the interplay of the recoil and tidal capture, 
the fitted $\alpha$ and $\beta$ values are typically slightly larger 
at a given separation and evolutionary stage. 
As discussed in Section \ref{s:binary_kick_model}, this arises from the fact that recoil produces a small number of systems just on the brink of tidal capture. 

The most crucial takeaway from this exercise is that 
the orbital distribution functions of wide MS+MS, WD+MS, and WD+WD binaries are inevitably different. 
Even if, for some reason, neither recoil nor tidal capture occurred in real binary systems, 
we would expect to observe the imprint of adiabatic orbital expansion 
in the separation-dependent value of $\alpha$. 
However, if either or both of these effects occur, 
we predict that they would each leave a characteristic signature in the wide binary eccentricity distribution. 
Recoil excites orbital eccentricities on average, typically increasing $\alpha$ at a given separation; 
tidal capture produces a turnover or cutoff for $e \gtrsim 0.9$, 
producing positive $\beta$ values. 

\begin{figure*}
    \centering
    \includegraphics[width=\textwidth]{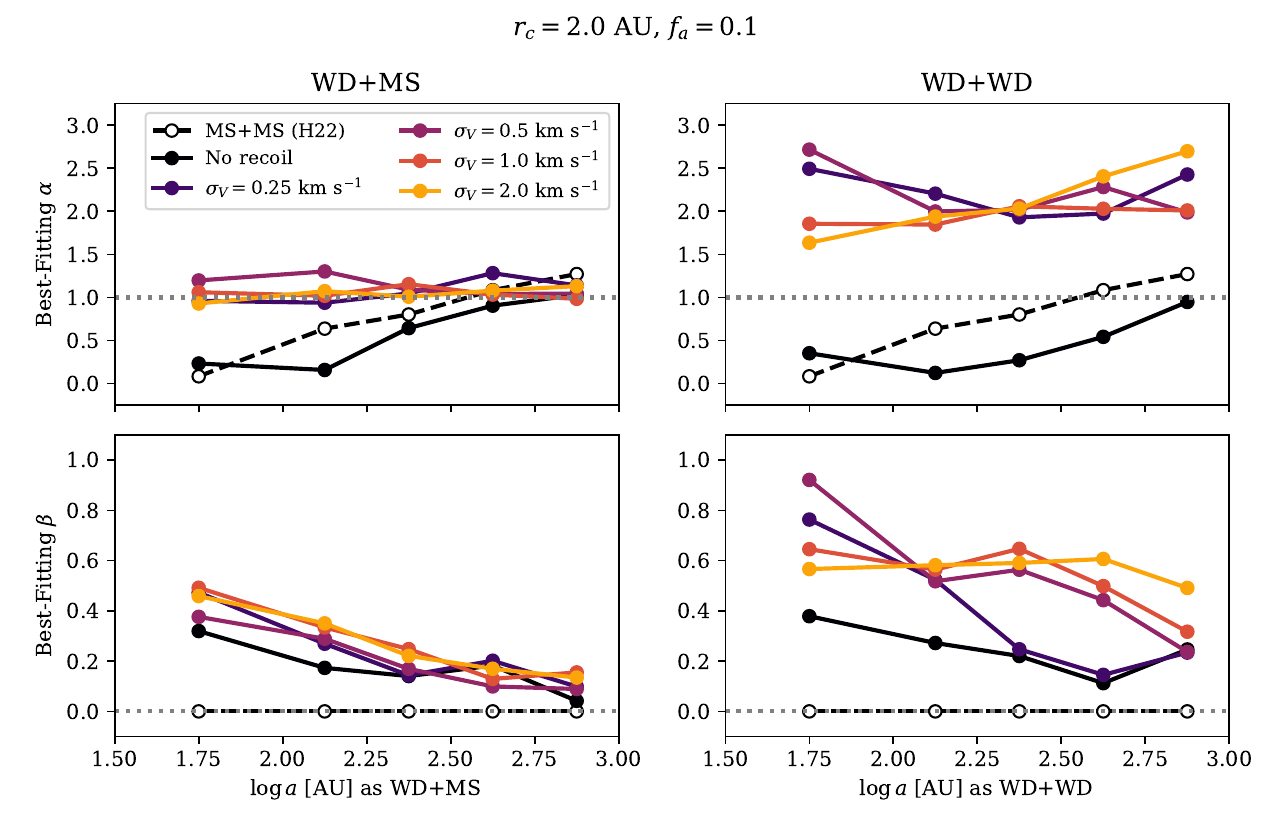}
    \caption{Best-fitting $\alpha$ and $\beta$ values for the eccentricity distributions of simulated binaries as a function of semi-major axis. 
    The left-hand panels show systems during the WD+MS stage, 
    while the right-hand panels show them in the WD+WD stage. 
    For the filled points connected by solid lines, 
    different colors correspond to different $\sigma_{V}$ values, 
    with $r_{c} = 2.0 \AU$ and $f_{a} = 0.1$ in all cases; 
    larger $\sigma_{V}$ values correspond to brighter colors. 
    For comparison, the assumed properties of the initial (MS+MS) eccentricity distribution 
    are shown as empty black circles connected by dashed black lines. 
    The $\alpha = 1$ and $\beta = 0$ for a thermal distribution are also shown as grey dotted lines.}
    \label{fig:compare_alphabeta_varysigv}
\end{figure*}

\begin{figure*}
    \centering
    \includegraphics[width=\textwidth]{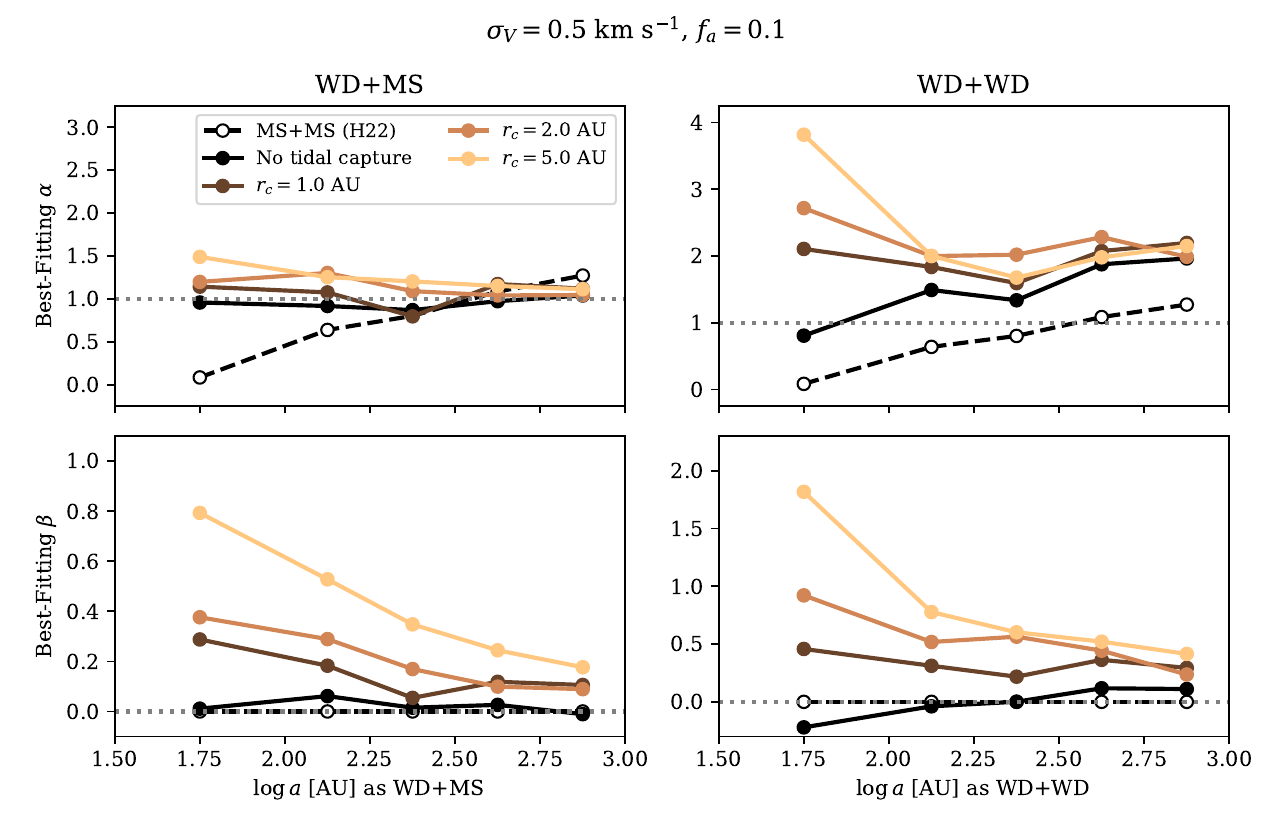}
    \caption{Similar to Fig.\ \ref{fig:compare_alphabeta_varysigv}, 
    but varying the tidal capture radius $r_{c}$ with $\sigma_{V} = 0.5 \, \kms$ held constant. 
    Larger $r_{c}$ values correspond to lighter colors.}
    \label{fig:compare_alphabeta_varyrcol}
\end{figure*}

\begin{figure*}
    \centering
    \includegraphics[width=\textwidth]{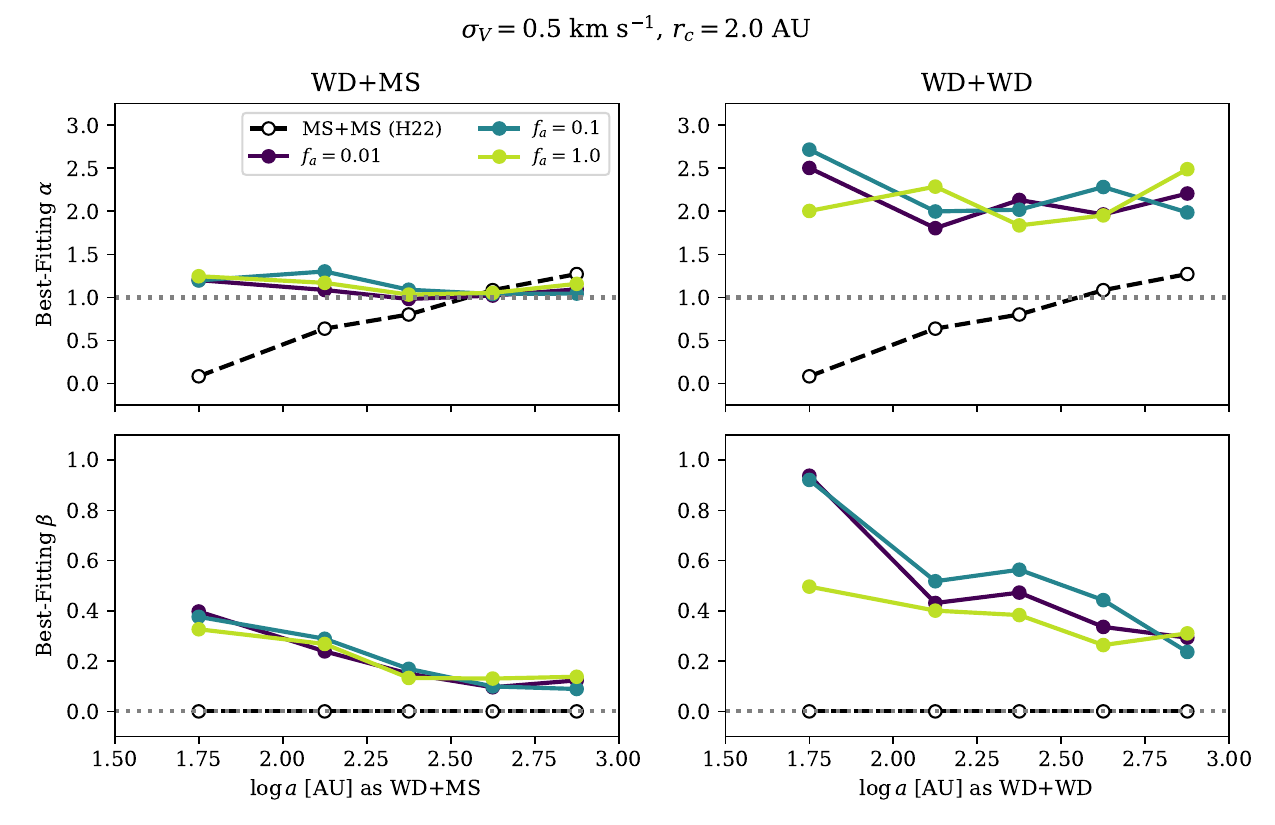}
    \caption{Similar to Figs.\ \ref{fig:compare_alphabeta_varysigv} and \ref{fig:compare_alphabeta_varyrcol}, 
    but varying the recoil timescale parameter $f_{a}$ with $\sigma_{V} = 0.5 \, \kms$ and $r_{c} = 2.0 \AU$ held constant. 
    Larger $f_{a}$ values correspond to brighter colors.}
    \label{fig:compare_alphabeta_varyfa}
\end{figure*}

\subsection{Population synthesis} \label{s:gaia_binaries:fullpopsynth}

Having established clearly how an idealized binary population evolves, 
we now conduct a parameter study of the evolution of more realistic stellar populations. 
We draw primary stellar masses at birth $m_{1i}$ from an initial mass function 
$\dif N/\dif m_{1i} \propto m_{1i}^{-2.3}$ on the interval $1.0 \leq m_{1i}/\MSol \leq 8.0$ \citep{Salpeter1955, Kroupa2001}. 
We obtain the secondary initial mass by drawing the binary mass ratio on the MS, $q = m_{2i} / m_{1i}$, 
from a flat distribution on the interval $0.08 \leq q \leq 0.95$.\footnote{
We did not consider the population of `twin' binaries, systems with $0.95 \leq q \leq 1$. 
Twin binaries display an excess of highly eccentric systems ($e \gtrsim 0.9$) 
compared to the overall wide binary population \citep{Hwang2022_twins}.} 
We generated initial orbital semi-major axes and eccentricities 
as described in the previous section. 

In these experiments, we varied the parameters $\sigma_{V}$, $r_{c}$, and $f_{a}$ 
in a controlled fashion 
covering combinations of the following values: 
$\sigma_{V} = \{ 0, 0.25, 0.5, 1.0, 2.0 \} \kms$, $f_{a} = \{ 0.01, 0.1, 1 \}$,\footnote{For 
$\sigma_{V} = 0$, varying $f_{a}$ is redundant. 
We therefore consider only $f_{a} = 1$ in those instances.} 
and $r_{c} = \{ 0, 1, 2, 5 \} \AU$. 
Since models with $\sigma_{V} = 0$ do not experience recoil, 
the parameter $f_{a}$ is not meaningful in these cases; 
hence, we only consider variations of $r_{c}$ when $\sigma_{V} = 0$. 
Each ensemble contained $10^{4}$ binary systems initially, 
fewer of which reach the WD+MS and WD+WD stages. 

For brevity's sake, we do not present detailed results for all combinations of parameters in our model grid. 
It suffices to say the population-level properties evolve similarly to those in the idealized systems in Set II (see Section \ref{s:gaia_binaries:setII}). 
In Figures \ref{fig:compare_alphabeta_varysigv}, \ref{fig:compare_alphabeta_varyrcol}, and \ref{fig:compare_alphabeta_varyfa}, 
we illustrate the effect of varying $\sigma_{V}$, $r_{c}$, and $f_{a}$ in a controlled fashion for a representative subset of the ensembles. 

Figure \ref{fig:compare_alphabeta_varysigv} shows the best-fitting $\alpha$ and $\beta$ values 
for all ensembles with $r_{c} = 2 \AU$ and $f_{a} = 0.1$ in order to gauge the effect of varying $\sigma_{V}$. 
Their values during the WD+MS and WD+WD stages are shown in the left- and right-hand panels, respectively. 
The most apparent differences we see are between the $\alpha$ values in ensembles with $\sigma_{V} = 0$ 
(no recoil, shown as black filled points connected by solid lines) and all others (color-coded by $\sigma_{V}$). 
For WD+MS systems, the $\alpha$ and $\beta$ values at a given $a$ mostly cluster together when $\sigma_{V} \geq 0.25 \,\kms$. 
However, these quantities exhibit greater variation for WD+WD systems: 
a close examination of the right-hand panels of Fig.\ \ref{fig:compare_alphabeta_varysigv} suggests 
that measuring both $\alpha$ and $\beta$ to $\pm 0.1$ for WD+WD systems across the range $1.5 \leq \log(a/{\rm AU}) \leq 3.0$ 
could constrain $\sigma_{V}$ to within a factor of $2$. 
This could be further improved by demanding consistency with the demographics of much wider ($a \gtrsim 10^{3} \AU$) WD+MS and WD+WD binaries \citep{ElBadry2018}.

Figure \ref{fig:compare_alphabeta_varyrcol} shows the effects of varying $r_{c}$ at constant $\sigma_{V}= 1 \, \kms$ and $f_{a} = 0.1$. 
As expected, we find that $\beta$ increases substantially with $r_{c}$ for both WD+MS and WD+WD systems. 
Values decrease to a moderate-to-large degree as a function of $a$ in cases with $r_{c} \neq 0$. 
Since $\sigma_{V}$ is held fixed in this sequence of models, we would not na\"{i}vely expect $\alpha$ to vary much as $r_{c}$ changes. 
This expectation is mostly borne out for WD+MS systems, where we find $1 \lesssim \alpha \lesssim 1.5$ in all cases. 
On the other hand, the WD+WD systems exhibit a surprisingly strong correlation between $r_{c}$ and $\alpha$ at $a \lesssim 10^{2} \AU$. 
Thus, while the eccentricity distribution of WD+MS systems does not constrain $\sigma_{V}$ very well, 
it is much more sensitive to $r_{c}$ -- i.e., to the efficiency of tidal capture during the AGB stage. 
This highlights why it is important to distinguish between the WD+MS and WD+WD samples in \gaia's binary catalog 
when studying the post-MS evolution of wide binaries. 

Lastly, Figure \ref{fig:compare_alphabeta_varyfa} shows the effects of varying $f_{a}$ at constant $\sigma_{V} = 0.5 \, \kms$ and $r_{c} = 2 \AU$. 
This exercise assess the impact of uncertainties in the timing of WD recoil. 
To reiterate, larger values of $f_{a}$ imply that the AGB star has lost a smaller fraction of its mass when directional mass loss begins; 
this translates to a smaller average torque on the binary due to recoil, since the orbit will be physically smaller when recoil sets in. 
It is certainly plausible that $f_{a}$ could vary within a real stellar population, 
due to the complexity of wind-driven stellar mass loss. 
As it turns out, though, this quantity has a mostly negligible effect on the eccentricity distribution, with no consistent trend across $0.01 \leq f_{a} \leq 1$. 
Thus, the eccentricity distribution of binaries inside $\sim 10^{3} \AU$ 
is not very sensitive to the timing of a proto-WD's recoil acceleration, provided the adiabatic approximation remain valid.  

\subsection{Tidal capture events}

Thus far we have focused on systems that avoid tidal capture and hence survive as wide binaries. 
We now examine the subset of systems that undergo tidal capture. 
Although predicting the subsequent evolution and fates of these systems is beyond the scope of our current study, 
it is of interest to understand how the probability of tidal capture depends on $\sigma_{V}$, $r_{c}$, and $f_{a}$ in our model. 
This quantity relates to the potential rate at which tidal capture events occur in the Galaxy. 

We define the quantity $F_{c1}(a)$ as the fraction of MS+MS systems with semi-major axis $a$ 
that undergo tidal capture in the process of evolving into a WD+MS system. 
Similarly, we define $F_{c2}(a)$ as the fraction of WD+MS systems whose {\it initial} semi-major axes were $a$ 
that undergo capture while evolving into a WD+WD system. 
These quantities give the probability of tidal capture events where the secondary star is a MS star or a WD, respectively, 
as a function of the binary's semi-major axis at birth. 

\begin{figure*}
    \centering
    \includegraphics[width=\linewidth]{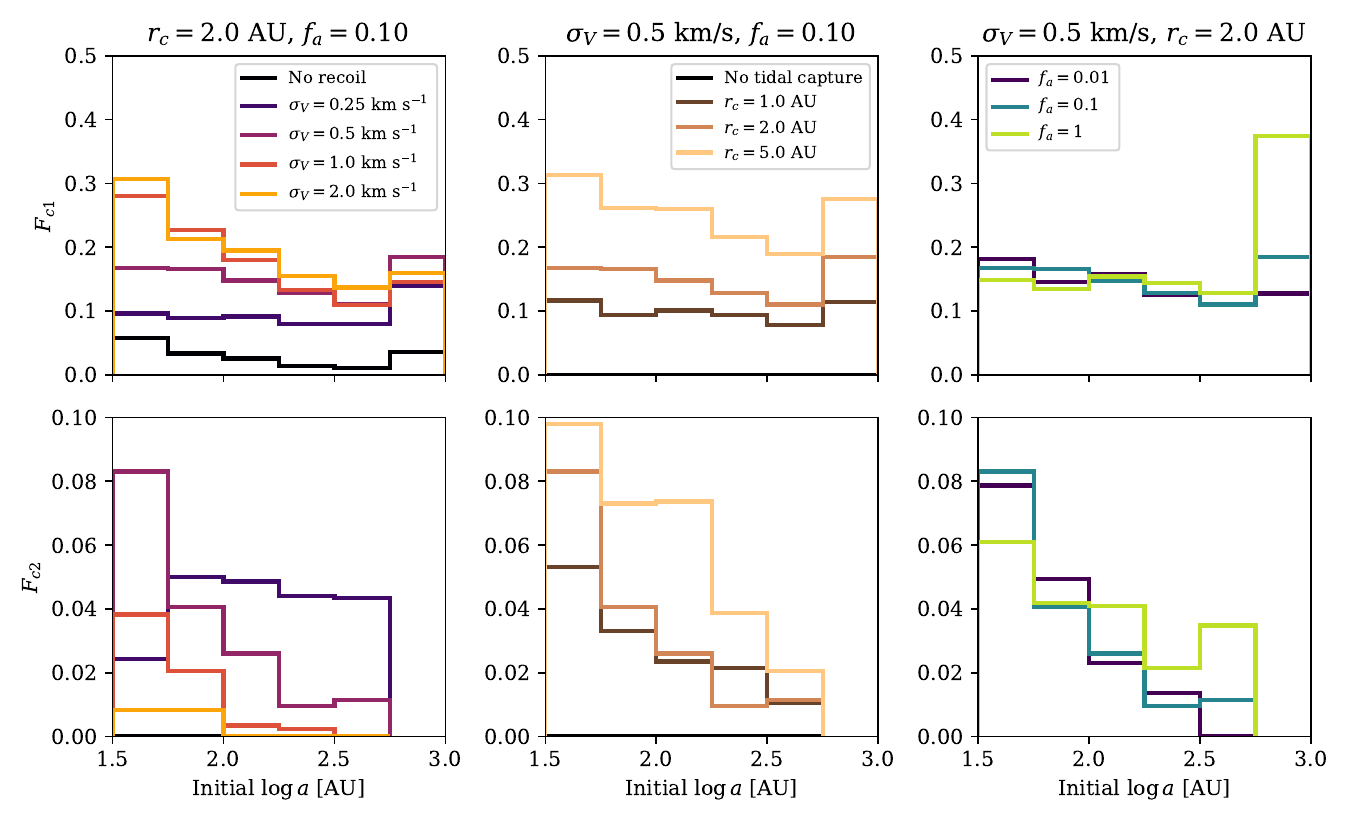}
    \caption{Stair plots showing the probability of tidal capture events during the primary AGB stage ($F_{c1}$, upper row) 
    and secondary AGB stage ($F_{c2}$, lower) as functions of initial semi-major axis. 
    Panels in the left, center, and right columns show the effects of varying $\sigma_{V}$, $r_{c}$, and $f_{a}$, respectively, 
    using the same color schemes as Figs.\ \ref{fig:compare_alphabeta_varysigv} through \ref{fig:compare_alphabeta_varyfa}.}
    \label{fig:tidal_capture_fractions}
\end{figure*}

We measure $F_{c1}$ and $F_{c2}$ in our population synthesis experiments within six bins of equal logarithmic width in $a$. 
We do this by counting the systems with initial separations in each bin 
that undergo tidal capture in the primary and secondary AGB stages; 
this count excludes systems that reach $a > 10^{3} \AU$ through orbital expansion before tidal capture occurs, 
since the adiabatic approximation breaks down in those cases. 
This cut mostly affects systems formed in the outermost bin.
We then divide by the total number of WD+MS or WD+WD progenitor systems at each stage. 

Figure \ref{fig:tidal_capture_fractions} shows $F_{c1}$ and $F_{c2}$ as functions of initial semi-major axis for the experiments of Section \ref{s:gaia_binaries:fullpopsynth}. 
The left, center, and right panels show how $\sigma_{V}$, $r_{c}$, and $f_{a}$ affect these quantities, respectively. 
For higher $\sigma_{V}$ values, i.e.\ greater degrees of secular phase-mixing, $F_{c1}$ increases at a given $a$, 
reaching $\approx 0.3$ among the closest systems for $\sigma_{V} \gtrsim 1 \, \kms$. 
However, this corresponds to a decrease in $F_{c2}$, since fewer WD+MS systems 
survive on orbits vulnerable to tidal capture during the secondary AGB phase. 
We see the largest $F_{c2} \approx 0.08$ for $\sigma_{V} = 0.5 \kms$. 
Meanwhile, increasing $r_{c}$ (center panel) monotonically increases $F_{c1}$ and $F_{c2}$. 
Again, the maximum values in this sequence of experiments are $F_{c1} \approx 0.3$ and $F_{c2} \approx 0.1$ in the closest separation bin. 
Finally, the probability of tidal capture is mostly insensitive to $f_{a}$, 
with the exception of $F_{c1}$ for $\log{a} \in [2.75,3.0]$; 
in that case, $F_{c1}$ increases from $\approx 0.12$ at $f_{a} = 0.01$ to $\approx 0.4$ at $f_{a} = 1$. 
This happens because when $f_{a}$ is close to unity, systems are able to have their eccentricities excited 
before their orbits expand beyond the adiabatic limit of $a \sim 10^{3} \AU$. 

It is notable that all the results shown in Figure \ref{fig:tidal_capture_fractions} feature an uptick in $F_{c1}$ 
(but not $F_{c2}$) in the widest separation bin, $\log a \in [2.75,3.0]$. 
This is a consequence of the initial eccentricity distribution in this range, 
a superthermal distribution $f_{e} \propto e^{1.2}$, 
which increases the proportion of systems with initially small periapsis separations. 
This highlights one of the more counterintuitive results of our work: 
that wide binaries can have a substantial chance of post-MS tidal interactions or mergers, 
thanks to the interplay of superthermal eccentricities and directional mass loss. 

\section{Discussion} \label{s:discussion}

\subsection{Constraining AGB mass loss physics with \gaia\ astrometry}

We have shown that, by virtue of the nature of mass loss and tidal interactions on the AGB, 
the eccentricity distribution of wide binaries inevitably differs between MS+MS, WD+MS, and WD+WD binaries 
(see also \citealt{HwangZakamska2025}). 
Current and future astrometric datasets offer the chance to probe this distribution 
and thereby the underlying physics of these processes. 

Our model for post-MS binary evolution relies on several simplifying assumptions. 
The most important of these is that WD recoil is well described by a small, adiabatic acceleration in a fixed direction. 
Other assumptions could plausibly have been made: 
For instance, the instantaneous acceleration could vary stochastically, 
in which case the orbital evolution might be better described as diffusion in phase space. 
This could be the case if recoil occurs through the ejection of a series of asymmetric mass shells with different orientations, 
perhaps corresponding to different thermal pulses.
In any given system, however, this hypothetical diffusive evolution results in a well-defined net velocity boost over a certain time interval; 
our model might be viewed as crudely estimating the final state of the orbit in this scenario.

Additionally, the direction of the recoil acceleration could be correlated with the stellar axis of rotation, rather than isotropically distributed as we have assumed. 
The rotation axis may, in turn, be correlated with the orientation of the orbit. 
This could lead to markedly different predicted eccentricity distributions and tidal capture rates, 
since the orbital geometry sets the amplitude of the orbital modulation. 
A future work aimed at studying these variant scenarios would require fundamentally different methods from ours, 
such as a direct integration of the Newtonian equations of motion (see Appendix \ref{app:validate}) 
coupled to a specific stellar evolution model.

\subsection{Rates and byproducts of tidal capture events}

We have measured the probability of highly eccentric close encounters ($a[1-e] \lesssim 5 \AU$) in binary systems 
in a scenario where WD progenitor stars receive $\sim \kms$ recoil velocities with an isotropic distribution. 
We have assumed that these close encounters result in ``tidal capture'' events, 
which may range in character from tidal circularization on a relatively close orbit ($\sim 1 \mbox{--} 10 \AU$) 
to the onset of a CE phase with a high eccentricity. 
Whatever their outcome may be, we can estimate the total rate of tidal captures as
\begin{equation} \label{eq:rate_tidal_capture}
    \Gamma_{c1} \sim F_{b} \langle F_{c1} \rangle \Gamma_{\rm WD} = 0.1 \yr^{-1} \left( \frac{\Gamma_{\rm WD}}{1 \yr^{-1}} \frac{F_{b}}{0.5} \frac{\langle F_{c1} \rangle}{0.2} \right),
\end{equation}
where $\Gamma_{\rm WD}$ is the Galactic rate of WD formation, $F_{b}$ is the wide binary fraction of WD progenitors, 
and $\langle F_{c1} \rangle$ is the weighted average of $F_{c1}$ over the binary period distribution. 
We may estimate $\Gamma_{c2}$, the rate of captures among AGB stars with WD companions, by substituting $F_{c2}$ for $F_{c1}$; 
it is typically about an order of magnitude smaller than $\Gamma_{c1}$. 

Our results suggests the rate of tidal capture events in initially wide binaries may be substantial 
regardless of the uncertainties and approximations inherent in our model for wide binary evolution. 
In the remainder of this section, we discuss two possible outcomes (which are not mutually exclusive): 
slow red transients arising from partial or total ejection of the AGB envelope, 
and the formation of a circularized WD+MS or WD+WD binary after a common-envelope phase. 

\subsubsection{Rate of slow red transients}

The prospect of inducing stellar collisions via tidal capture during the late AGB stage 
has clear implications for optical and infrared time-domain astronomy. 
AGB stars are already bright and variable at these wavelengths, 
making them identifiable {\it en masse} in other galaxies. 
The engulfment of a MS or WD companion would trigger the expansion and possible ejection of the AGB envelope on a dynamical timescale, 
producing a moderately bright (peak $M_{r} \sim -6$), long-duration ($\sim 1 \yr$), dust-obscured eruption 
\citep[cf.][]{Tylenda2013_OGLE2002BLG360, MacLeod+2022, OConnorBildstenCantielloLai2023, Karambelkar+2025_AGBtransient}. 
Using the Galactic event rate for tidal capture events estimated above (equation \ref{eq:rate_tidal_capture}), 
and adopting a number density of $0.02 \, {\rm Mpc}^{-3}$ for Milky Way--like galaxies, 
we predict a rate density of $\sim 0.002 \yr^{-1} \, {\rm Mpc}^{-3}$ for these events in the local universe. 
This implies total rates of $\sim 0.5 \yr^{-1}$ out to a distance of $\sim 4 \, {\rm Mpc}$ (i.e., within the Local Group) 
and $\sim 70 \yr^{-1}$ with $\sim 20 \, {\rm Mpc}$ (out to the Virgo Cluster). 

Indeed, a population of slow transients associated with red giant progenitors has begun to emerge in wide-field optical and near-infrared surveys. 
Only a few individual events have been positively identified thus far, starting with the Galactic transient OGLE-2002-BLG-360 \citep{Tylenda2013_OGLE2002BLG360}. 
\citet{Karambelkar+2025_AGBtransient} reported the discovery of WNTR23bzdiq, 
a similar event in M31, and the first with a confirmed AGB progenitor. 
While the rate of these events has not been precisely determined, the order of magnitude appears consistent 
with equation (\ref{eq:rate_tidal_capture})'s crude estimate \citep[e.g.,][]{Kochanek2014_mergerrate, Karambelkar2023_LRNeRates}. 
We therefore suggest that eccentric wide binaries be considered as a potential source population for these events. 
Further study of highly eccentric CE events is needed to identify any distinguishing features they may have, 
particularly relative to classical CE events in close, circular binary systems. 
We anticipate that as infrared time-domain astronomy matures, the rate and characteristics of transients towards AGB progenitors will become clearer. 

\subsubsection{Byproducts of tidal capture}

What is the long-term fate of a tidally captured system? 
There are two main possibilities, depending on whether the stars enter a CE phase following tidal capture. 
This bifurcation is most likely decided by the efficiency of tidal dissipation in the AGB star's envelope during the close periapsis passages ($r_{p} \sim r_{c}$) by the companion. 

When the periapsis separation approaches $r_{c} \sim \text{a few } R_{1}$, the companion raises strong tides on the AGB star's envelope; 
the dissipation of those tides creates friction, tending to shrink and circularize the orbit. 
We may roughly gauge whether the system enters a CE phase by comparing the eccentricity excitation timescale $T_{\rm Stark}$ (equation \ref{eq:def_TStark}) with the tidal circularization timescale ($T_{\rm circ}$, see below). 
We estimate the latter using the theory of weak tidal friction \citep[e.g.,][]{Hut1981_weakfriction}; 
this provides a good description of the tidal evolution of eccentric binaries containing giant stars with relatively compact companions, 
where the dissipation arises from convective turbulent viscosity in the envelope \citep[e.g.,][]{VerbuntPhinney1995, VickLai2020}. 
Taking the limit $e \to 1$ and neglecting the rotation of the AGB star, the orbital circularization timescale is given by:
\begin{align} \label{eq:Tcirc}
    T_{\rm circ}&\simeq \frac{m_{1} r_{p}^{13/2} a^{3/2}}{2 G M m_{2} R_{1}^{5} k_{21} \tau_{1}}
\end{align}
where $k_{2,1}$ is the giant's tidal Love number and $\tau_{1}$ is the tidal lag-time. 
For definiteness, we set $k_{2,1} \tau_{1} \approx 0.024 (R_{1}^{3} / G m_{1})^{1/2}$ following calculations for representative red giant models by \citet{VickLai2020}. 
Equation (\ref{eq:Tcirc}) then yields 
\begin{align}
    T_{\rm circ} &\approx 1.5 \times 10^{5} \yr \left( \frac{r_{p}}{2 R_{1}} \right)^{13/2} \left( \frac{a}{100 \AU} \right)^{3/2} \nonumber \\
    & \hspace{0.5cm} \times \left( \frac{m_{1}}{\MSol} \right)^{3/2} \left( \frac{m_{2}}{\MSol} \right)^{-1} \left( \frac{M}{2 \MSol} \right)^{-1}.
\end{align}

\begin{figure}
    \centering
    \includegraphics[width=\columnwidth]{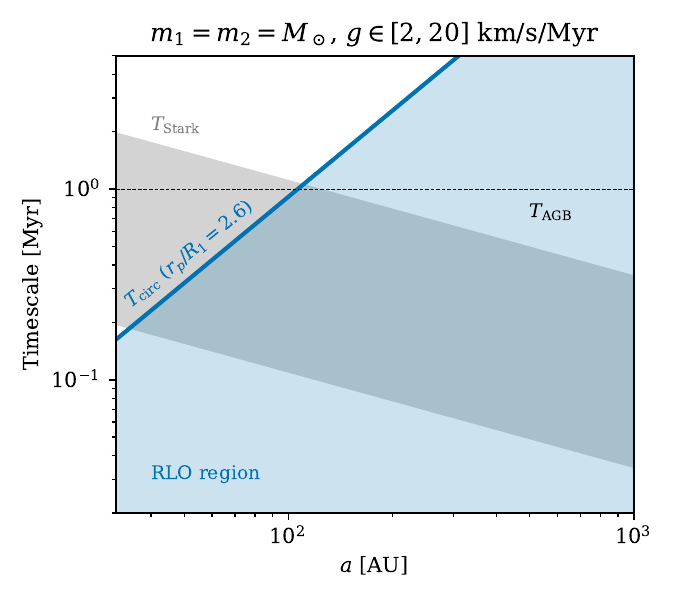}
    \caption{Timescales for eccentricity excitation and circularization in a highly eccentric binary system with a $1 \MSol$ AGB primary and a $1 \MSol$ point-like companion. 
    The grey shaded region indicates a range of eccentricity excitation timescales $T_{\rm Stark}$ (equation \ref{eq:def_TStark}) for recoil accelerations $g \in [2,20] \, \kms \Myr^{-1}$. 
    The thick blue line shows the circularization timescale $T_{\rm circ}$ at $r_{p}/R_{1} = 2.6$. 
    In the shaded blue region beneath this curve, the AGB star overflows its Roche lobe at periapsis. 
    The thin, horizontal line marks the typical lifetime of an AGB star.}
    \label{fig:stark_circ_times}
\end{figure}

Figure \ref{fig:stark_circ_times} illustrates the comparison of $T_{\rm Stark}$ and $T_{\rm circ}$ for a system with $m_{1} = m_{2} = \MSol$. 
The light grey region shows $T_{\rm Stark}$ as a function of semi-major axis for a range of recoil accelerations $g \in [2,20] \, \kms \Myr^{-1}$. 
The thick blue line shows $T_{\rm circ}$ for $r_{p}/R_{1} = 2.6$, 
below which (light blue shading) the AGB component overflows its Roche lobe at periapsis, 
most likely entering a CE phase.  
We see that $T_{\rm Stark} \lesssim T_{\rm circ}$ for binaries with $a \gtrsim 100 \AU$. 
In this regime, recoil can drive the binary towards RLO faster than tidal friction can circularize the orbit. 
For $a \lesssim 100 \AU$, there is a range of accelerations for which circularization may dominate. 
Hence, we expect tidal capture events in binaries with $a \gtrsim 100 \AU$ 
to drive the stars into a CE phase with a high eccentricity. 
Closer binaries may instead undergo partial circularization while avoiding a CE phase. 
However, we note that the competition between eccentricity pumping by recoil and damping by tidal friction 
may vary wildly, even within a given system: 
The radius of a late-stage AGB star varies on a timescale $\lesssim 10^{4} \yr$ due to thermal pulses, 
and the circularization timescale is highly sensitive to the ratio $r_{p}/R_{1}$.  

\paragraph{Eccentric barium stars}

Tidally captured binaries that avoid a CE phase are potential progenitors of eccentric barium (Ba) stars. 
These form when a MS star accretes from the wind or envelope of a thermally pulsing AGB companion, 
thereby becoming enriched in Ba and other $s$-process elements \citep[e.g.,][]{McClure1984}. 
A persistent difficulty in the theory of Ba star evolution is an inability to match their observed period--eccentricity distribution with standard binary evolution models, 
owing mainly to the high efficiency of tidal circularization at periods $P \lesssim 3000 \, {\rm d}$ on the AGB \citep[e.g.,][]{Krynski+2025}. 
\citet{Izzard+2010} studied the potential role of WD recoil in exciting the orbital eccentricities of Ba stars, 
obtaining a rough match to the observations with an assumed $\Delta V \sim 4 \, \kms$. 
However, certain details invite further attention: 
First, \citeauthor{Izzard+2010} used to calculate the orbital evolution from recoil 
in a period regime where orbit-averaging is more appropriate ($P \lesssim 10^{5}$ days). 
Second, the large $\Delta V$ required to match the period--eccentricity distribution for Ba stars in their calculations 
is no longer supported by \gaia\ astrometry, at least not for the majority of WDs \citep[][]{ElBadry2018, HwangZakamska2025}. 
Third, mass transfer in eccentric binaries may, under certain circumstances, act to increase eccentricities rather than damp them \citep[e.g.,][]{Rocha+2025}. 
In light of these considerations, the potential for Ba star formation from tidally captured wide binaries should be explored in a future study. 

\paragraph{AU-scale WD binaries}

If the binary enters a CE phase before circularizing, strong gravitational and hydrodynamical drag forces will dissipate the companion's orbital energy, rapidly reducing the orbital period and eccentricity. 
Since the envelope of a late AGB star is diffuse and tenuously bound, the engulfment of a stellar-mass companion can readily disperse it.  
We can estimate the companion's final orbital semi-major axis $a_{\rm post}$ using the energy formalism for CE evolution \citep[e.g.,][]{Webbink1984}. 
The post-CE semi-major axis may be estimated as
\begin{subequations} \label{eq:apost_CE}
\begin{align}
    a_{\rm post} &\simeq \alpha_{\rm CE} \frac{G m_{\rm 1c} m_{2}}{2 |E_{\rm b1}|} \\ 
    &= \frac{\alpha_{\rm CE} \kappa R_{1}}{2} \left( \frac{m_{\rm 1c} m_{2}}{m_{1}^{2}} \right) \\
    &= 0.75 \AU \left( \frac{R_{1}}{1 \AU} \right) \left( \frac{\alpha_{\rm CE}}{0.3} \frac{\kappa}{10} \right) \nonumber \\
    & \hspace{1.5cm} \times \left( \frac{m_{\rm 1c}}{0.5 \MSol} \right) \left( \frac{m_{2}}{\MSol} \right) \left( \frac{m_{1}}{\MSol} \right)^{-2},
\end{align}
\end{subequations}
where the binding energy of $m_{1}$'s envelope is given by $|E_{\rm b1}| = G m_{1}^{2}/ \kappa R_{1}$ 
and where $\alpha_{\rm CE}$ is the CE efficiency parameter. 
We set $\alpha_{\rm CE} = 0.3$ based on empirical constraints 
derived from the properties of post-CE WD binaries (e.g., \citealt{Zorotovic2010, ZorotovicSchreiber2022, ScherbakFuller2023}; but see \citealt{Yamaguchi+2025}). 
We adopt fiducial values for the AGB star by consulting Table 1 of \citet{OConnorBildstenCantielloLai2023}, 
which lists these quantities for a selection of 1D models for typical solar-mass RGB and AGB stars. 
Given the natural variation of stellar properties between different systems, one can find reasonable combinations of parameters 
giving $a_{\rm post}$ a few times greater or smaller than the reference value in equation (\ref{eq:apost_CE}). 
It is more difficult to estimate the final orbital eccentricity; 
however, smoothed-particle hydrodynamics simulations of eccentric CE phases in relatively close binaries by \citet{GlanzPerets2021} yielded post-CE eccentricities as high as $0.4$. 
In our scenario, the companion's initial period and eccentricity would both be somewhat greater than that study considered; 
hence, even higher final eccentricities seem plausible, depending on how many times the companion passes through the envelope. 
Further study is required to understand the range of possible outcomes for highly eccentric CE events. 

Thus, we predict that CE evolution initiated by tidal capture of wide binary companions during the late AGB would produce circular or moderately eccentric WD+MS and WD+WD systems at AU orbital scales. 
The WD components would universally have C/O cores, having passed undisturbed through most of the AGB phase. 
These characteristics are suggestively similar to the population of WD+MS binary systems 
discovered in the \gaia\ DR3 astrometric dataset by \citet{Shahaf2024_wdbinarytriage} \citep[see also][]{Yamaguchi+2024, Yamaguchi+2025}, 
which exhibit typical orbital periods $10^{2} \lesssim P/{\rm day} \lesssim 10^{3}$; 
eccentricities $0 \leq e \lesssim 0.8$ (mostly $\lesssim 0.3)$; MS component masses of $\sim 0.4 \mbox{--} 1.2 \MSol$; and WD component masses mostly above $0.55 \MSol$. 
As \citet{Shahaf2024_wdbinarytriage} discussed, standard binary evolution models struggle to reproduce this population. 
Our crude estimate above suggests that recoil-induced CE events in wide binaries can potentially account for their properties and occurrence rate. 
In our population synthesis calculations, we found that a substantial fraction of wide MS+MS binaries -- 
up to $\sim 30\%$, depending on separation and uncertain model parameters (see Fig.\ \ref{fig:tidal_capture_fractions}) -- 
could enter a high-eccentricity CE phase during the late AGB stage. 

Equation (\ref{eq:apost_CE}) implies correlations between the properties of the WD and MS components of AU-scale binaries. 
All else being equal, we would expect systems with more massive MS companions (higher $m_{2}$) to have longer orbital periods on average, 
since those companions can more easily disperse an AGB star's envelope 
and thus may undergo less inspiral during the CE phase. 
On the other hand, more massive WD components (higher $m_{\rm 1c}$) have more massive progenitors (higher $m_{1}$) 
with greater binding energies; 
thus, at a fixed $m_{2}$, we would expect orbital periods to {\it decrease} with increasing $m_{\rm 1c}$. 
All of these predictions presuppose that the energy formalism suitably describes the outcomes of high-eccentricity CE events 
and that the efficiency factor $\alpha_{\rm CE}$ is similar to the value inferred from WD systems with close M-dwarf, brown dwarf, and WD companions 
\citep[e.g.,][]{Zorotovic2010, ZorotovicSchreiber2022, ScherbakFuller2023}.

Finally, we predict the existence of a population of AU-scale WD+WD binaries analogous to the WD+MS systems of \citet{Shahaf2024_wdbinarytriage}. 
These would have undergone an eccentric CE phase during the secondary star's AGB phase, 
i.e.\ that of the initially less-massive star in the system. 
We obtain the relative numbers of tidal capture events during the primary and secondary AGB stages in our models, 
finding $F_{c2}/F_{c1} \lesssim 0.5$ (see Fig.\ \ref{fig:tidal_capture_fractions}). 
Extrapolating from the $\sim 3000$ WD+MS candidates reported by \citet{Shahaf2024_wdbinarytriage} 
implies that there may be up to $\sim 1500$ AU-scale WD+WD binaries within $1 \, {\rm kpc}$ of the Sun. 
Based on \gaia\ EDR3 data, \citet{Korol+2022} reported a moderate excess of double WD binaries at AU separations, 
where standard binary population synthesis models predict a `gap' in the separation distribution carved out by CE evolution. 
Further study of this putative excess in subsequent \gaia\ data releases can evalaute whether a tidal-capture origin is feasible; 
constraints on the orbital eccentricities of these systems would be especially informative. 

\section{Conclusion} \label{s:conclusion}

In this study, we have considered the dynamical evolution of wide binary systems 
with primary masses of $\sim 1 \mbox{--} 8 \MSol$. 
We have accounted for (i) the eccentricity distribution as a function of separation as measured by \gaia\ astrometry among MS binaries, 
(ii) orbital modulation due to WD recoil from asymmetric AGB winds, 
and (iii) the possibility of strong tidal interactions or stellar collisions induced by (ii). 
Our main conclusions are as follows:
\begin{enumerate}
    \item If WD progenitor stars typically experience $\sim 1 \, \kms$ of recoil during the late AGB stage, 
    the direction of which is chosen at random from a spherically isotropic distribution, 
    then the eccentricity of a binary with separation $\lesssim 10^{3} \AU$ will be modified significantly. 
    However, the system will remain bound, since the acceleration occurs in an adiabatic, orbit-averaged fashion. 

    \item Binaries driven to high eccentricities ($\gtrsim 0.95$) by recoil may undergo tidal capture at periapsis separations of a few AU if the recoil occurs while the AGB star retains an extended envelope. 
    The probability of tidal capture prior to the WD+MS (WD+WD) stage may be as high as $30\%$ ($10\%$), depending on the critical radius for tidal capture and the typical $\Delta V$ for recoil. 
    
    \item Based on simple population synthesis experiments, we predict that the distribution of orbital eccentricity as a function of orbital separation differs significantly between the MS+MS, WD+MS, and WD+WD samples in \gaia's wide binary catalogue. 
    The nature of the differences record the effects of both WD recoil and tidal capture in the wide binary population. 

    \item We suggest that wide binaries that undergo tidal capture during the primary AGB phase may enter a common-envelope phase, 
    becoming the progenitors of eccentric WD+MS binaries with $\sim 1 \AU$ separations \citep{Shahaf2024_wdbinarytriage}. 
    We predict the existence of a smaller, analogous population of WD+WD binaries with similar periods and eccentricities, resulting from tidal capture during the secondary AGB phase \citep[cf.][]{Korol+2022}. 
    The coalescence of such binaries and ejection of the envelope may be observable as slow optical/infrared transients with AGB progenitors \citep[e.g.,][]{Tylenda2013_OGLE2002BLG360, Karambelkar+2025_AGBtransient}. 
\end{enumerate}
Future works can broaden the scope of these results by performing self-consistent orbital integrations of individual binaries with a realistic prescription for stellar mass loss and radius evolution; 
this will also require accounting for external perturbations, such as Galactic tides and stellar flybys, that we have been able to neglect in this work. 

WD recoil has significant implications for the dynamical evolution of other astrophysical systems, such as globular clusters and planetary systems. 
In globular clusters with velocity dispersions of a few $\kms$, WD recoil can provide a sustained source of dynamical heating, 
delaying the onset of core collapse \citep{Fregeau+2009}. 
Consequently, WD recoil may indirectly regulate the formation of low-mass X-ray binaries, millisecond pulsars, fast radio bursts, and other exotica 
via dynamical interactions in core-collapsed cluster environments \citep[e.g.,][]{SigurdssonPhinney1995, Ivanova2008_NScluster, Ye2019_MSPcluster, Kremer2023_frbcluster}. 
In planetary systems, meanwhile, the excitation of eccentricities and mutual inclinations by WD recoil 
may trigger dynamical instabilities promptly after the death of the host star \citep[e.g.,][]{Akiba+2024, Stephan+2024}. 
This would have major implications for the origin of metal pollution in WD atmospheres, 
as well as the architectures of surviving planetary systems and planetesimal reservoirs around WDs \citep[e.g.,][]{ParriottAlcock1998, Stone+2015, OConnorLiuLai2021, OConnorLaiSeligman2023}. 
Finally, the occurrence of WD recoil complicates the longstanding question of the fate of the Solar System. 
Previous studies of the stability of the outer planets during and after the Sun's post-MS evolution have only considered isotropic mass loss 
\citep[e.g.,][]{DuncanLissauer1998, Zink+2020_solarsystem}; 
these generally predict that the outer planets remain stable for at least a Hubble time after the death of the Sun. 
In a forthcoming work, we will revisit the problem taking WD recoil into account. 

Near the completion of this work, \citet{HwangZakamska2025} put forth an independent study along the same lines, 
arriving at several similar conclusions. 
Those authors have modeled the post-MS evolution of wide binaries 
in more or less the same way as we have, with one key difference: 
They have not taken recoil-induced tidal capture events into account. 
As we have shown, this is a crucial element of the story for binaries wider than $\sim 100 \AU$, 
due to the interplay of the initial eccentricity distribution 
with the large level-arm for secular torques from WD recoil. 
Indeed, \citet{HwangZakamska2025} report a dearth of WD--MS binaries with eccentricities greater than $0.94$ in the \gaia\ sample,
consistent with our population synthesis results when the tidal capture radius $r_{c}$ is finite. 

\section*{Acknowledgements}

C.E.O.\ thanks Tatsuya Akiba, Katie Breivik, Kyle Kremer, Yoram Lithwick, Ann-Marie Madigan, Fred Rasio, Cheyanne Shariat, and Alexander Stephan for helpful discussions during the development of this work. 
C.E.O.\ was supported by a CIERA Postdoctoral Fellowship. 
This work used computing resources provided by Northwestern University and the Center for Interdisciplinary Exploration and Research in Astrophysics (CIERA). 
This research was supported in part through the computational resources and staff contributions 
provided for the Quest high performance computing facility at Northwestern University 
which is jointly supported by the Office of the Provost, 
the Office for Research, and Northwestern University Information Technology.

\software{{\tt REBOUND} \citep{ReinLiu2012_rebound}, {\tt REBOUNDx} \citep{Tamayo2020_reboundx}, {\tt scipy} \citep{Virtanen2020_scipy}}

\appendix

\section{General solution of the Stark problem with adiabatic mass variation} \label{app:stark_general_solution}

For a general initial orbit with Keplerian elements ($e_{0}, I_{0}, \omega_{0}, \Omega_{0}$), 
the initial vector elements are
\begin{subequations}
\begin{align}
    \vect{j}(0) &= \left( 1 - e_{0}^{2} \right)^{1/2} \left[ \zhat \cos{I_{0}} + \sin{I_{0}} \left( \xhat \sin{\Omega_{0}} - \yhat \cos{\Omega_{0}} \right) \right], \\
    \vect{e}(0) &= e_{0} \left[ \zhat \sin{I_{0}} \sin{\omega_{0}} + \xhat \left( \cos{\Omega_{0}} \cos{\omega_{0}} - \sin{\Omega_{0}} \sin{\omega_{0}} \cos{I_{0}} \right) + \yhat \left( \sin{\Omega_{0}} \cos{\omega_{0}} + \cos{\Omega_{0}} \sin{\omega_{0}} \cos{I_{0}} \right) \right].
\end{align}
\end{subequations}
For a constant Stark frequency $\gamma$, the general solution is
\begin{subequations} \label{eq:stark_solutions}
\begin{align}
    \vect{j}(t) &= \xhat \left[ \left( 1 - e_{0}^{2} \right)^{1/2} \sin{I_{0}} \sin{\Omega_{0}} \cos{(\gamma t)} + e_{0} \left( \cos{I_{0}} \cos{\Omega_{0}} \sin{\omega_{0}} + \sin{\Omega_{0}} \cos{\omega_{0}} \right) \sin{(\gamma t)} \right] \nonumber \\
    & \hspace{0.5cm} 
    + \yhat \left[ - \left( 1 - e_{0}^{2} \right)^{1/2} \sin{I_{0}} \cos{\Omega_{0}} \cos{(\gamma t)} + e_{0} \left( \cos{I_{0}} \sin{\Omega_{0}} \sin{\omega_{0}} - \cos{\Omega_{0}} \cos{\omega_{0}} \right) \sin{(\gamma t)} \right] \nonumber \\
    & \hspace{0.5cm} + \zhat \left( 1 - e_{0}^{2} \right)^{1/2} \cos{I_{0}}, \\
    \vect{e}(t) &= \xhat \left[ - \left( 1 - e_{0}^{2} \right)^{1/2} \sin{I_{0}} \cos{\Omega_{0}} \sin{(\gamma t)} - e_{0} \left( \cos{I_{0}} \sin{\Omega_{0}} \sin{\omega_{0}} - \cos{\Omega_{0}} \cos{\omega_{0}} \right) \cos{(\gamma t)} \right] \nonumber \\
    & \hspace{0.5cm} + \yhat \left[ - \left( 1 - e_{0}^{2} \right)^{1/2} \sin{I_{0}} \sin{\Omega_{0}} \sin{(\gamma t)} + e_{0} \left( \cos{I_{0}} \cos{\Omega_{0}} \sin{\omega_{0}} + \sin{\Omega_{0}} \cos{\omega_{0}} \right) \cos{(\gamma t)} \right] \nonumber \\
    & \hspace{0.5cm} + \zhat \, e_{0} \sin{I_{0}} \sin{\omega_{0}} .
\end{align}
\end{subequations}
If $\gamma$ is itself a slowly varying function of time, 
then equations (\ref{eq:stark_solutions}) may be modified to give a correct solution 
by replacing the product $\gamma t$ with the quantity 
\begin{equation}
    \phi(t) \equiv \int_{0}^{t} \gamma(t') \, \dif t'. 
\end{equation}
For the special case of adiabatic, directional mass loss with a constant effective exhaust velocity $\vect{w}_{k}$, 
we may effect the following change of variables: 
\begin{align}
    \phi(t) = \int_{0}^{t} \gamma(t') \, \dif t' &= \int_{0}^{t} \frac{3}{2} \left[ \frac{a(t')}{GM(t')} \right]^{1/2} g(t') \, \dif t' \nonumber \\
    &= - \int_{m_{ka}}^{m_{k}(t)} \frac{3}{2} \left[ \frac{a(m_{k}')}{GM(m_{k}')} \right]^{1/2} \frac{g_{k}(m_{k}') \, \dif m_{k}'}{\dot{m}_{k}(m_{k}')} \nonumber \\
    &= \frac{3}{2} \frac{V_{k}}{\ln(m_{ka}/m_{kf})} \int_{m_{ka}}^{m_{k}(t)} \left[ \frac{a(m_{k}')}{GM(m_{k}')} \right]^{1/2} \frac{\dif m_{k}'}{m_{k}'}.
\end{align}
By equation (\ref{eq:a(t)_secular}), we have
\begin{equation}
    a(m_{k}') = a_{i} \left[ \frac{m_{ki} + m_{3-k}}{M(m_{k}')} \right],
\end{equation}
with $M(m_{k}') = m_{k}' + m_{3-k}$ for $k=1,2$ as appropriate. 
With this substitution, 
we obtain
\begin{align}
    \phi(t)
    = \frac{3}{2} \frac{V_{k}}{\ln(m_{ka}/m_{kf})} \left( \frac{a_{i}}{G m_{3-k}} \right)^{1/2} \left( 1 + \frac{m_{ki}}{m_{3-k}} \right)^{1/2} \left[ \ln\left( \frac{m_{ka} + m_{3-k}}{m_{k}(t) + m_{3-k}} \right) - \ln\left( \frac{m_{ka}}{m_{k}(t)} \right) \right].
\end{align}
For $m_{ki} \simeq m_{ka} \simeq m_{kf}$, 
we recover the correct result for the Stark problem with constant masses and a weak external field. 

\section{Validation of the secular model} \label{app:validate}

To validate the analytical model presented in the main text in Section \ref{s:binary_kick_model} and Appendix \ref{app:stark_general_solution}, 
we compare its predictions with direct integrations of the Newtonian equations of motion. 
To facilitate this comparison and to provide a simple, reliable tool for future studies, 
we implemented rocket-like mass loss in {\tt REBOUND} \citep{ReinLiu2012_rebound},
an open-source software package for N-body simulations in widespread use, 
via the extension package {\tt REBOUNDx} \citep{Tamayo2020_reboundx}. 
In this Appendix, we present our implementation of a new {\tt REBOUNDx} operator, called {\tt rocket} (Section \ref{app:validate:rocket}); 
and use {\tt rocket} to test the secular model. 

\subsection{Rocket-like mass loss in {\tt REBOUNDx}} \label{app:validate:rocket}

{\tt REBOUNDx} currently supplies an operator {\tt modify\_mass} that causes particles in a {\tt REBOUND} simulation 
to undergo exponential mass growth or mass loss with a user-specified e-folding timescale \citep{Kostov2016_reboundx}. 
This operator does not include a recoil acceleration by default. 
We have implemented a new operator, {\tt rocket}, designed to emulate non-relativistic rocket-like mass loss. 
It can also be used to model accretion by a non-relativistic object moving through a medium, 
as in Bondi--Hoyle--Lyttleton accretion. 
This operator implements mass loss or growth at a constant rate, rather than exponential mass changes as in {\tt modify\_mass}. 

The {\tt rocket} operator accepts up to four parameters for each particle in the simulation: 
{\tt mdot}, the rate of change of the particle's mass; 
{\tt veff}, the effective exhaust speed; 
{\tt theta}, the polar angle (i.e., co-latitude) of the kick with respect to the positive $z$-axis; 
and {\tt phi}, the azimuthal angle of the kick in the $xy$-plane measured counterclockwise from the $x$-axis. 
All parameters are set equal to zero by default; 
any parameter can be adjusted at each timestep during an integration, if desired. 

At the beginning of each timestep, {\tt rocket} updates the mass and velocity of each mass-losing particle in the simulation according to its specified parameters, 
then transforms the system into the new center-of-momentum frame. 
This mirrors the approach taken by {\tt modify\_mass}. 
The source code for {\tt rocket} and instructions for incorporating it into {\tt REBOUNDx} are freely available for download.\footnote{\doi{10.5281/zenodo.17087919}} 

\subsection{Worked examples}

The following {\tt PYTHON} code snippet shows the necessary steps to include rocket-like mass loss in a {\tt REBOUND}/{\tt REBOUNDx} simulation
of a $2 \MSol + 1 \MSol$ binary system 
with initial $a = 300 \AU$ and $e = 0.75$. 
The more massive object (particle ``A'' below) evolves into a $0.6 \MSol$ WD 
over an elapsed time of $0.1 \Myr$, 
receiving a total velocity boost of $1 \, \kms$
towards $(\theta,\phi) = (87.5^{\circ}, 0^{\circ})$. 

\begin{verbatim}
import rebound, reboundx
import numpy as np

PI = np.pi
km = 1./1.496e8
sec = 1./3.156e7

# create simulation object
sim = rebound.Simulation()
sim.units = ("AU","yr","Msun") # use Sun-Earth units

mAi = 2. # Msun
mAf = 0.6 # Msun
mB = 1. # Msun
vk = 1. * km/sec
T = 1.5e6 # years

# add rocket-effect operator
rebx = reboundx.Extras(sim)
rocket = rebx.load_operator("rocket")
sim.add_operator(rocket)

# add particles
sim.add(m=mAi,hash="A")
sim.add(m=mB,primary="A",a=100.,e=0.75,hash="B")
sim.move_to_com()

# set rocket operator parameters for particle A
mdot = (mAf-mAi)/T
veff = vk/np.log(mAi/mAf)
sim.particles["A"].params["mdot"] = mdot
sim.particles["A"].params["veff"] = veff
sim.particles["A"].params["theta"] = 87.5 * (PI/180)
sim.particles["A"].params["phi"] = 0.
\end{verbatim}

\begin{figure}
    \centering
    \includegraphics[width=0.5\textwidth]{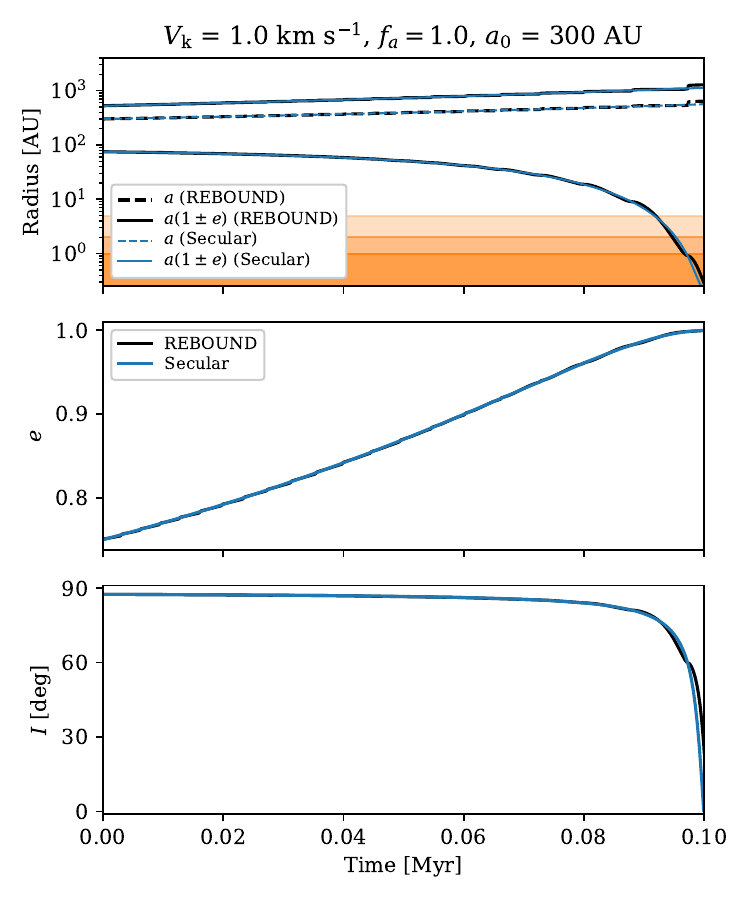}
    \caption{Comparison between the analytical secular solution (blue curves) and direct integration of the equations of motion in {\tt REBOUND}/{\tt REBOUNDx} (black). 
    The binary's initial conditions are as follows: $m_{\rm A} = 2 m_{\rm B} = 2 \MSol$, 
    $a = 300 \AU$, $e = 0.75$, $I = 87.5 \deg$, $\Omega = \omega = 0$. 
    The primary star's mass ($m_{\rm A}$) is reduced to $0.6 \MSol$ over $0.1 \Myr$, 
    receiving a $1.0 \, \kms$ velocity boost from recoil with $f_{a} = 1$. 
    {\it Top panel:} Evolution of semi-major axis (dashed curves) 
    and the periapsis and apoapsis separations (solid curves). 
    The light, medium, and dark orange shaded regions mark the fiducial 
    tidal capture radii used in Section \ref{s:gaia_binaries}, $r_{c} = \{ 1.0, 2.0, 5.0 \} \AU$. 
    {\it Middle:} Eccentricity evolution. 
    {\it Bottom:} Inclination evolution.} 
    \label{fig:rocket_test_3e2}
\end{figure}

Figure \ref{fig:rocket_test_3e2} shows the results of the integration alongside the secular solution presented in Section \ref{s:binary_kick_model} and Appendix \ref{app:stark_general_solution} for the same initial conditions. 
The direct integration and the secular solution are in good agreement, 
since the orbital period is much less than the Stark timescale (equation \ref{eq:def_TStark}). 

\begin{figure}
    \centering
    \includegraphics[width=0.5\textwidth]{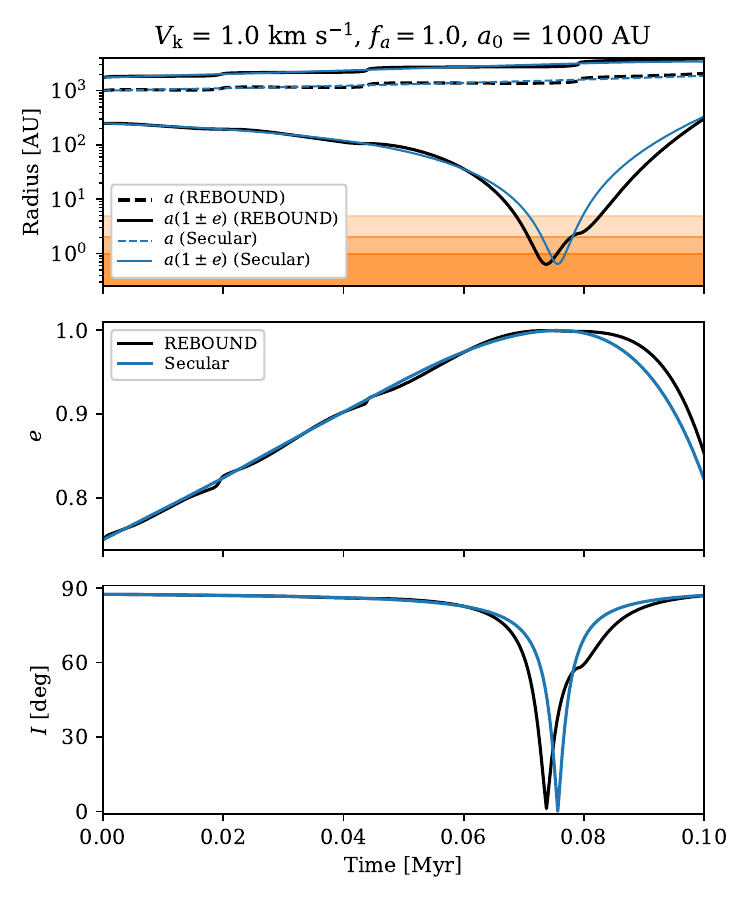}
    \caption{The same as Fig.\ \ref{fig:rocket_test_3e2}, but with initial $a = 1000 \AU$.}
    \label{fig:rocket_test_1e3}
\end{figure}

\begin{figure*}
    \centering
    \includegraphics[width=0.5\textwidth]{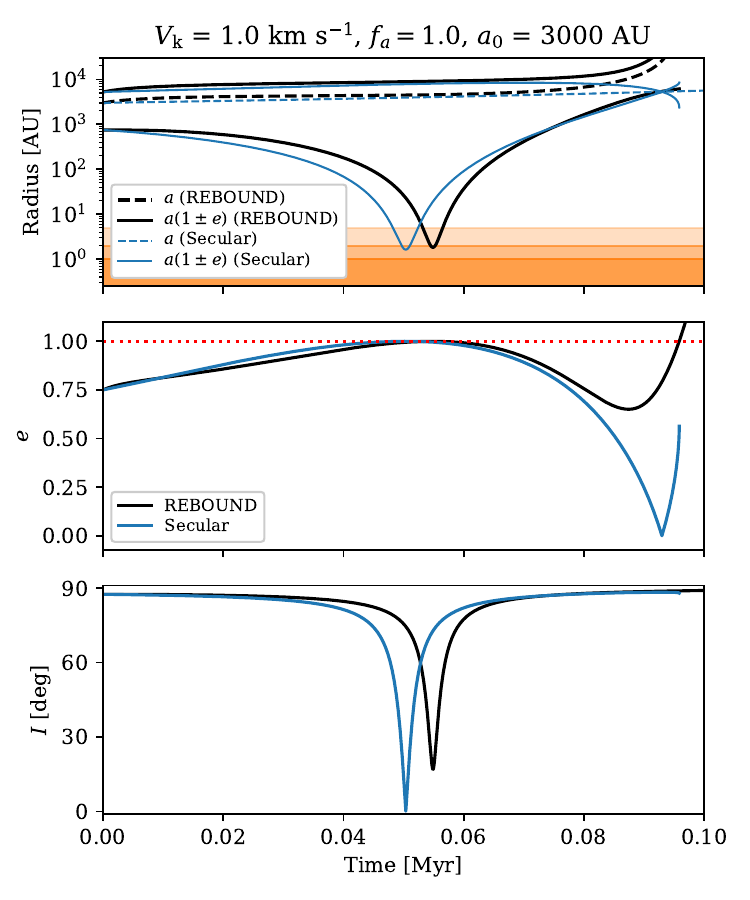}
    \caption{The same as Figs.\ \ref{fig:rocket_test_3e2} and \ref{fig:rocket_test_1e3}, 
    but with initial $a = 3000 \AU$. 
    In the middle panel, the horizontal dotted red line shows $e = 1$; 
    the point where the black curve crosses this line marks where recoil unbinds the system in the direct integration.}
    \label{fig:rocket_test_3e3}
\end{figure*}

We repeat this exercise for systems with initial $a = 1000 \AU$ and $a = 3000 \AU$, 
keeping all other parameters and initial conditions the same. 
Figures \ref{fig:rocket_test_1e3} and \ref{fig:rocket_test_3e3} compare the {\tt REBOUND} outputs to the secular solution. 
For initial $a = 1000 \AU$, modest deviations from the secular solution 
appear as the orbit expands, since the Stark timescale becomes comparable to the orbital period. 
For initial $a = 3000 \AU$, however, the secular solution breaks down entirely, 
the recoil being strong enough to dissociate the binary at late times. 

In each figure, we highlight the regions $r_{p} \leq r_{c}$ where, 
in our population synthesis experiments, 
a binary system undergoes tidal capture. 
We see that in each case, 
the secular solution predicts the minimum $r_{p}$ value achieved by the system quite accurately, 
even when other aspects of the solution break down. 
This suggests that the tidal capture fractions calculated in Section \ref{s:gaia_binaries} (Fig.\ \ref{fig:tidal_capture_fractions}) 
are fairly robust to deviations from the secular solution in a real system. 

\bibliography{ms_kick1_aastex7_arxiv}
\bibliographystyle{aasjournalv7}

\end{document}